\documentclass[preprint,showpacs,showkeys]{revtex4}

\usepackage{amssymb, amsmath, amsfonts, latexsym, verbatim, mathrsfs}

\usepackage{graphicx}

\usepackage{latexsym,amssymb,amsmath,amsfonts, amsthm, xcolor}

\newtheorem{theorem}{Theorem}
\newtheorem{itlemma}{Lemma}[section]
\newtheorem{itproposition}[itlemma]{Proposition}
\newtheorem{itcorollary}[itlemma]{Corollary}
\newtheorem{itremark}[itlemma]{Remark}
\newtheorem{itremarks}[itlemma]{Remarks}
\newtheorem{itdefinition}{Definition}
\newtheorem{itexample}[itlemma]{Example}
\newtheorem{conjecture}{Conjecture}

\newenvironment{remark}{\begin{itremark}\rm}{\end{itremark}} 

\newenvironment{proposition}{\begin{itproposition}\rm}{\end{itproposition}}
\newenvironment{definition}{\begin{itdefinition}\rm}{\end{itdefinition}}

\begin{document}

\title{Minimum time control of a pair of two-level quantum systems with opposite drifts}
\thanks{Work supported by ARO MURI under Grant W911NF-11-1-0268}

\author{Raffaele Romano}
\email{rromano@iastate.edu}
\affiliation{Department of Mathematics, Iowa State University, 50011 Ames, IA, USA \,}

\author{Domenico D'Alessandro}
\email{daless@iastate.edu}
\affiliation{Department of Mathematics, Iowa State University, 50011 Ames, IA, USA}


\begin{abstract}
In this paper we solve two equivalent time optimal control problems. On one hand, we design
the control field to implement in minimum time the SWAP (or equivalent) operator on a two-level
system, assuming that it  interacts with an additional, uncontrollable, two-level system. On
the other hand, we synthesize the SWAP operator simultaneously, in minimum time, on a pair of two-level
systems subject to opposite drifts. We assume that it is possible to perform three independent
control actions, and that the total control strength is bounded. These controls
either affect the dynamics of the target system, under the first perspective, or, simultaneously,
the dynamics of both systems, in the second view.
We obtain our results by using techniques of geometric control theory on Lie groups. In particular,
we apply the Pontryagin Maximum Principle, and provide a complete characterization of singular and
non-singular extremals. Our analysis shows that the problem can be formulated as the motion of a
material point in a central force, a well known system in classical mechanics. Although we focus
on obtaining the SWAP operator, many of the ideas and techniques developed in this work apply to
the time optimal implementation of an arbitrary unitary operator.

\end{abstract}

\pacs{02.30.Yy, 03.65.Aa, 03.67.-a}

\keywords{Optimal control, SU(2), Quantum dynamics}

\maketitle


\section{Introduction}

\subsection{Statement of the problem}

In this work we consider the controlled dynamics of a pair of interacting two-level quantum systems, where
the purpose of the control action is the generation in minimum time of a desired unitary operation
on one of these systems, which in the following will be called the {\it target system} ($T$). The other
system represents an unavoidable disturbance or {\it environment} ($E$). The dynamics of the whole system is
given by the Schr\"{o}dinger operator equation
\begin{equation}\label{schr}
    \dot{X} (t) = -i H(u) X (t), \qquad X (0) = I \otimes I,
\end{equation}
where $X (t)$ describes the evolution of the density matrix of the composite system from its initial
to its final configuration, $\rho (0) \rightarrow \rho (t) = X (t) \rho (0) X^{\dagger} (t)$, and $I$
is the identity operator for two-level systems. The initial
state is assumed to be factorized, $\rho(0) = \rho_E(0) \otimes \rho_T(0)$, and
$H(u)$ is the Hamiltonian of the system, given by
\begin{equation}\label{hami}
    H(u) = I \otimes (u_x S_x + u_y S_y + u_z S_z) + 2 \omega_0 S_z \otimes S_z.
\end{equation}
The first term in $H (u)$ contains three possibly time varying control actions $u_k = u_k (t)$, $k = x, y, z$, with
$L_2$ norm bounded in strength, $u_x^2 + u_y^2 + u_z^2 \leqslant \gamma^2$. These control actions are associated
with the generators $S_k = \frac{1}{2} \sigma_k$, $k = x, y, z$, where $\sigma_k$ are the Pauli matrices
\begin{equation}\label{paul}
    \sigma_x = \left(
                 \begin{array}{cc}
                   0 & 1 \\
                   1 & 0 \\
                 \end{array}
               \right), \quad
    \sigma_y = \left(
                 \begin{array}{cc}
                   0 & -i \\
                   i & 0 \\
                 \end{array}
               \right), \quad
    \sigma_z = \left(
                 \begin{array}{cc}
                   1 & 0 \\
                   0 & -1 \\
                 \end{array}
               \right),
\end{equation}
with commuting relations $[\sigma_k, \sigma_l] = 2 i \sigma_m$, and $(k, l, m)$ is a cyclic permutation
of $(x, y, z)$\footnote{Although we assume that three independent control actions can affect the system's
dynamics, we will find that our results are valid also in the case with only two independent control actions entering
the dynamics, when we limit our attention to $X_f = X_{SWAP}$.}.
The second term in (\ref{hami}) represents an Ising interaction between the two systems; we
assume $\omega_0 \ne 0$, otherwise the problem reduces to that of controlling an isolated two-level
system with bounded control, which has been solved in previous work~\cite{AD,RR}. Without loss
of generality we can take $\omega_0 > 0$, as we shall shortly see.
We neglect the free evolution of the environmental two-level system. It is meant that terms on the right of
the tensor product are associated to the target system. Because of the block structure of the Hamiltonian
(\ref{hami}) and of the initial state $X (0)$, the evolution operator has a block structure at any time $t$,
\begin{equation}\label{diag}
    X (t) = {\rm diag} (X_1 (t), X_2 (t)) = P_1 \otimes X_1 (t) + P_2 \otimes X_2 (t),
\end{equation}
where the projectors $P_1$, $P_2$ are given by
\begin{equation}\label{proj}
    P_1 = \frac{1}{2} (I + \sigma_z), \qquad P_2 = \frac{1}{2} (I - \sigma_z),
\end{equation}
and the dynamics of the two blocks is given by
\begin{eqnarray}\label{schr2}
    \dot{X}_1 (t) &=& -i (\omega_0 S_z + u_x S_x + u_y S_y + u_z S_z) X_1 (t), \quad X_1 (0) = I, \nonumber \\
    \dot{X}_2 (t) &=& -i (-\omega_0 S_z + u_x S_x + u_y S_y + u_z S_z) X_2 (t), \quad X_2 (0) = I.
\end{eqnarray}
The dynamics of the target system is obtained by taking the partial trace of $\rho (t)$ over the degrees of
freedom of the environmental system; since ${\rm Tr} P_i P_j = \delta_{ij}$, we find
\begin{equation}\label{redu}
    \rho_T (t) = {\rm Tr}_E (\rho (t)) = {\rm Tr}_E \big(P_1 \rho_E (0)\big) X_1(t) \rho_T (0) X_1^{\dagger}(t) +
    {\rm Tr}_E \big(P_2 \rho_E (0)\big) X_2(t) \rho_T (0) X_2^{\dagger}(t).
\end{equation}
Our task is to find the minimum time $t_f$ such that this evolution reproduces a given unitary operator $X_f \in SU (2)$
acting on the target system,
\begin{equation}\label{fina}
\rho_T (t_f) = X_f \rho_T (0) X_f^{\dagger}.
\end{equation}
In principle, the initial state of the environmental qubit $\rho_E (0)$ is unknown. Therefore, Eq.s (\ref{redu})
and (\ref{fina}) must hold for every $\rho_E (0)$. A necessary and sufficient condition for this is that $X_1
(t_f) = \pm X_f$ and $X_2 (t_f) = \pm X_f$, with all possible combinations of signs. In view of the
evolution equations (\ref{schr2}), we conclude that the problem of optimally steering a two-level system in interaction
with a second two level system is equivalent to the problem of optimally driving two independent two-level systems,
subject to opposite drifts, through the same control actions. Moreover, we see that our choice $\omega_0 > 0$ is
not restrictive, since the case $\omega_0 < 0$ amounts to interchanging the two systems $1$ and $2$.

\subsection{Motivation}

The application to the study of quantum mechanical systems of methods of geometric control theory,
and in particular optimal control theory, is a longstanding practice. These methods are particularly
suitable in applications to the synthesis of specific operations in quantum information processing~\cite{nielsen},
the control of nuclear spins in nuclear magnetic resonance~\cite{levitt}, or the optimization of specific reactions
in atomic or molecular physics.

In this context, two-level systems are often the building blocks of larger structures. Therefore,
it is of great relevance to understand to what extent it is possible to manipulate them, especially
when they are immersed in a dissipative environment (see~\cite{wu,boscain,wenin,carlini,kirillova,garon,hegerfeldt,aiello,russell}
and references therein for some applications of optimal control theory to quantum systems). The standard approach for
their control is given by dynamical decoupling techniques~\cite{LidarReview}, which, however, usually require unbounded
control actions. Whenever a bound on the controls cannot be neglected, optimal control represents an alternative tool to dynamical
decoupling.

A dissipative environment is usually modeled by a thermal bath with infinitely many degrees of
freedom. However, for some specific systems, it is meaningful to model the external world by
means of a finite-dimensional system. A noticeable example is provided by the study of NV centers
in diamonds, which are currently of great interest for applications in quantum information
technology (see, e.g.~\cite{Slava1,Slava2}). The control task is to drive a central spin independently
of the environmental spins, whose coupling strengths to the central spin generally depend on their distances
from it. The problem considered in this work is the simplest instance of this situation where only one environmental
spin exists. See~\cite{Burgarth} for a numerical analysis of the dependence of the optimal time of transition
on the number of environmental spins.

In all these scenarios it is desirable that a given state transfer occurs in {\it minimum time}. This is
both to increase the efficiency of quantum operations which typically require a cascade of elementary
gates and to minimize the effect of an (un-modeled) environment.

Besides applications in quantum mechanics, the optimal control of $SU(2)$ operations is
strictly related to the optimal control on $SO(3)$, being $SU(2)$ the double cover of $SO(3)$.
Since $SO(3)$ is the group modeling the attitude of a rigid body, our analysis is
also of general interest in the study of classical mechanical systems.

\subsection{Plan of the paper}

We are able to derive analytical expressions for the time optimal control strategies, and the corresponding
optimal times, for target operations equivalent to the SWAP operation, in the terms described in
Proposition~\ref{rem1}, and for a specific range of values of $\omega_0$ and $\gamma$. Nonetheless,
whenever appropriate, we present partial result valid for general target operations $X_f$.

In Section~\ref{sec2} we discuss the Lie algebraic structure of the problem, and briefly review the Pontryagin
Maximum Principle (PMP) of optimal control theory~\cite{pontryagin,Flerish} in the context of control on Lie
groups. This fundamental result  provides the theoretical framework for our investigation. The controls
satisfying the PMP are candidate optimal and they are called {\it extremals}, and they are classified in
{\it nonsingular} and {\it singular} extremals. Usually, the procedure for solving an optimal control problem
by means of the PMP consists of several steps, which in our specific case are given by: (i) solution of the
dynamical equations derived from the PMP; (ii) determination of the extremals; (iii) solution of the system
dynamics for these control strategies, and computation of the corresponding time of transition to the desired
target operator; (iv) determination of the optimal strategy by comparison of the different extremal strategies.
In general, some of these steps cannot be carried out analytically, and one has to resort to numerical computations.
In the present case, numerical analysis and the structure of the problem will suggest a {\it conjecture}, which,
when adopted, allows an analytical result of the optimal control problem when $X_f = X_{SWAP}$.

We shall present a  complete investigation of the aforementioned points (i) - (iv) for the system of interest here.
This process is complicated by the presence of singular arcs, which require a separate analysis (see~\cite{boscain2,wu2,lapert}
for applications of optimal control theory to quantum mechanical systems where singular solutions exist).
In Section~\ref{sec3} we show that, in both cases of nonsingular and singular extremals, the basic equations
derived by applying the PMP can be interpreted as the motion of a fictitious material point subject to a
velocity-dependent force. If we choose the final target operator as $X_f = X_{SWAP}$, this force becomes
conservative, and the system is equivalent to an integrable classical central force problem. This fact provides
a parallel between this optimal control problem and the (deeply investigated) analysis of the motion of a material
point in a central force~\cite{broucke}.

At this point, we find convenient to limit our attention to the case $X_f = X_{SWAP}$, and show
that the extremal trajectories correspond to motions with null angular momentum in the effective central-force
problem. This result is valid for both cases of nonsingular and singular extremals, which are separately
investigated in Sections~\ref{sec4} and \ref{sec5}. A complete treatment of singular extremals can be
performed for arbitrary values of $\gamma$ and $\omega_0$, but the investigation of nonsingular extremals is
increasingly difficult as $\frac{\gamma}{\omega_0}$ decreases, therefore we limit our attention to cases where
analytical solutions are possible. In Section~\ref{sec6}, by comparing the different types of
extremals previously found, we solve the optimal control problem for $X_f = X_{SWAP}$ in the
aforementioned range of values of $\omega_0$ and $\gamma$, and provide the optimal control strategies and the optimal
time. In Section~\ref{sec7} we discuss our findings, relate them to existing results,
and conclude.

For sake of clarity, the most technical results are reported in Appendices. Also, the most significant
results, which are frequently referred to throughout the paper, are presented as Propositions, and formally proved.
The reader not interested in the computational details could directly move to Section~\ref{sec6}, where the main results
are summarized. The investigation of cases other than $X_f = X_{SWAP}$ will be presented in a forthcoming paper, complementing
the analysis presented here.


\section{The Lie algebraic structure of the problem}\label{sec2}

As anticipated in the Introduction, the optimal control problem that we consider in this work is the following:
given the dynamical system (\ref{schr2}) and a target operation $X_f \in SU(2)$,
we want to find a control strategy $u_k (t)$, $k = x, y, z$, such that $X_1 (t_f) = \pm X_f$ and $X_2 (t_f) = \pm X_1 (t_f)$,
and $t_f$ is minimum~\footnote{The case of a final operation $X_f \in U(2)$ can be dealt with by adapting the results in $SU(2)$}.
First of all,  we prove that, for every $\phi \in \mathbb{R}$, the operator
\begin{equation}\label{Xtil}
    \hat{X}_f = e^{i \varphi S_z} X_f e^{-i \varphi S_z},
\end{equation}
is reached in the same optimal time as $X_f$.
\begin{proposition}\label{rem1}
Consider a pair of operators $X_f$ and $\hat{X}_f \in SU(2)$ which satisfy (\ref{Xtil}) for some $\varphi$. Assume that
there is a control strategy $u_k = u_k (t)$ ($k = x, y, z$), with $u_x^2 + u_y^2 + u_z^2 \leqslant \gamma^2$, such that
$X_1$ and $X_2$ are mapped to $X_f$ (or its negative) in time $t_f$. Then
there is a control strategy $\hat{u}_k = \hat{u}_k (t)$ ($k = x, y, z$), with $\hat{u}_x^2 + \hat{u}_y^2 + \hat{u}_z^2
\leqslant \gamma^2$, such that $X_1$ and $X_2$ are mapped to $\hat{X}_f$ (or its negative) in the same time $t_f$.
\end{proposition}
\noindent {\it Proof:} If we define $\hat{X}_1 = e^{i \varphi S_z} X_1 e^{-i \varphi S_z}$ and
$\hat{X}_2 = e^{i \varphi S_z} X_2 e^{-i \varphi S_z}$, we can write
\begin{eqnarray}\label{schr3bis}
    \dot{\hat{X}}_1 (t) &=& -i (\omega_0 S_z + \hat{u}_x S_x + \hat{u}_y S_y + \hat{u}_z S_z)
    \hat{X}_1 (t), \quad \hat{X}_1 (0) = I, \nonumber \\
    \dot{\hat{X}}_2 (t) &=& -i (-\omega_0 S_z + \hat{u}_x S_x + \hat{u}_y S_y + \hat{u}_z S_z)
    \hat{X}_2 (t), \quad \hat{X}_2 (0) = I,
\end{eqnarray}
where $\hat{u}_x = u_x \cos{\varphi} + u_y \sin{\varphi}$, $\hat{u}_y = - u_x \sin{\varphi} + u_y \cos{\varphi}$, and
$\hat{u}_z = u_z$. These controls satisfy the required constraint since $u_x^2 + u_y^2 + u_z^2 = \hat{u}_x^2
+ \hat{u}_y^2 + \hat{u}_z^2$, therefore they represent an admissible control strategy. Moreover, the dynamics
(\ref{schr3bis}) is formally the same as (\ref{schr2}), therefore the control strategy $\hat{u}_k = \hat{u}_k (t)$
generates the desired transformation $X_1 (t_f) = \pm X_f$ and $X_2 (t_f) = \pm X_1 (t_f)$. \\

This result is a consequence of a symmetry of the problem, defined by the transformations $\hat{\sigma}_k =
e^{i \varphi S_z} \sigma_k e^{-i \varphi S_z}$ for $k = x, y$, and $\hat{\sigma}_z = \sigma_z$. This change of
representation of the Pauli matrices does not affect the system, since it merely corresponds to a redefinition
of the controls, but it leaves invariant the equations of motion (\ref{schr2}). The particular form of the
controls in the dynamics is crucial for the existence of this symmetry, which is preserved even if there are
only two independent controls affecting $\sigma_x$ and $\sigma_y$ (that is, $u_z \equiv 0$).

Following Proposition~\ref{rem1}, we will consider equivalent two target operators which satisfy (\ref{Xtil})
for some $\varphi$. Therefore, it is sufficient to solve the optimal control problem for one of them to obtain
a complete description of the optimal control strategies and optimal time for all of them. We will rely on this
fact, when we will fully solve the case with $X_f = X_{SWAP}$ (or equivalent operators). In the representation
induced by (\ref{paul}), operators related by (\ref{Xtil}) differ by a phase in the off diagonal terms.

It is well known in quantum control theory~\cite{miko} that information on the structure of a control problem
can be obtained by analyzing the Lie algebra associated to the system. We define the block-diagonal matrices:
\begin{equation}\label{BC}
    B_k = \left(
            \begin{array}{cc}
              S_k & 0 \\
              0 & S_k \\
            \end{array}
          \right), \qquad
    C_k = \left(
            \begin{array}{cc}
              S_k & 0 \\
              0 & - S_k \\
            \end{array}
          \right),
\end{equation}
where $S_k = \frac{1}{2} \sigma_k$, and $0$ is the $2 \times 2$ null matrix, and $k = x, y, z$.
In terms of them, the evolution (\ref{schr}) takes the form
\begin{equation}\label{schr3}
    \dot{X} (t) = -i (\omega_0 C_z + u_x B_x + u_y B_y + u_z B_z) X (t), \quad X (0) = I \otimes I.
\end{equation}
Since we want to find the controls steering in minimum time $t_f$ the evolution operator to $X (t_f) =
{\rm diag} (X_f, X_f)$, or $X (t_f) = {\rm diag} (X_f, -X_f)$, or their negatives, where $X_f \in SU(2)$,
we first apply results on the {\it controllability} of the system, to answer the question of whether
this type of state transfer is possible. This is indeed the case since the Lie algebra generated by the
Hamiltonians $C_z$, $B_x$, $B_y$ and $B_z$ in (\ref{schr3}) is, in the considered representation, the Lie algebra
of block diagonal $4 \times 4$ matrices with arbitrary $2 \times 2$ blocks in $\mathfrak{su} (2)$. The set
of reachable operators is the associated Lie group of block diagonal $4 \times 4$ matrices with $2 \times 2$
blocks in $SU(2)$. This group contains the desired final conditions for any special unitary operator $X_f$.
Therefore, system (\ref{schr3}) is controllable on the Lie group of operators of the form ${\rm diag}(Y, Z)$,
where $Y, Z \in SU(2)$. This group is compact, semisimple, and isomorphic to $SU (2) \oplus SU(2)$.
The associated Lie algebra, isomorphic to $\mathfrak{su} (2) \oplus \mathfrak{su}(2)$, is given by
$\mathfrak{l} = \mathfrak{b} \oplus \mathfrak{c}$, where
\begin{equation}\label{liea}
    \mathfrak{b} = {\rm span} (B_x, B_y, B_z), \qquad \mathfrak{c} = {\rm span} (C_x, C_y, C_z)
\end{equation}
satisfy the commutation relations of a Cartan decomposition of $\mathfrak{l}$,
\begin{equation}\label{cart}
    [\mathfrak{b}, \mathfrak{b}] \subseteq \mathfrak b, \quad [\mathfrak{c}, \mathfrak{c}] \subseteq \mathfrak b, \quad
    [\mathfrak{b}, \mathfrak{c}] \subseteq \mathfrak c.
\end{equation}
More specifically, the commutation relations of the operators in (\ref{BC}) are given by
\begin{equation}\label{comm}
    [B_k, B_l] = i B_m, \quad [C_k, C_l] = i B_m, \quad [B_k, C_l] = i C_m,
\end{equation}
and $(k, l ,m)$ is a cyclic permutation of $(x, y, z)$. The controllability properties of the system along with
the fact that the set of possible values for the control is compact allows us to use the standard Filippov's
existence result for time optimal control problems (see, e.g.,~\cite{Flerish}) to conclude that the time optimal
control exists for every final condition in $SU(2) \oplus SU(2)$~\footnote{For the unbounded minimum time control
problem, one can find the optimal control in the sense of~\cite{Khaneja}. The form of the optimal control can be
significantly different for different types of bounds on the control such as bounds on the $L_1$ or $L_\infty$ norm.}.

To solve our problem, we shall apply the classical necessary conditions of optimality given by the {\it Pontryagin
Maximum Principle} (see, e.g.,~\cite{Flerish} and~\cite{miko2,Tesi} for application to quantum systems on Lie groups).
In order to state them in a form appropriate for our goals, we set up some definitions. Given $M \in \mathfrak{l}$
(usually called the {\it costate}), and $X$ a solution of (\ref{schr3}), we define the time-dependent coefficients
\begin{equation}\label{bc}
    b_k = i \langle M, X^{\dagger} B_k X \rangle, \quad c_k = i \langle M, X^{\dagger} C_k X \rangle,
\end{equation}
where $\langle A, B \rangle = {\rm Tr} (A B^{\dagger})$, and $k = x, y, z$.

\begin{definition}\label{def1}
The {\it Pontryagin Hamiltonian} is defined as
\begin{equation}\label{pont}
    H(M, X, v_x, v_y, v_z) = \omega_0 c_z + \sum_{k = x,y,z} v_k b_k.
\end{equation}
\end{definition}

\begin{theorem}\label{the1}
{\rm - Pontryagin Maximum Principle (PMP) -} Assume that the evolution of $X$ is described by (\ref{schr3}), and that
$u_x, u_y, u_z$ is the time optimal control strategy steering $X$ from the identity $I \otimes I$ to ${\rm diag}
(X_f, X_f)$ or  ${\rm diag} (X_f, -X_f)$. Denote by $\tilde{X} (t)$ the corresponding trajectory. Then there exists
an operator $\tilde{M} \in \mathfrak{l}$, $\tilde{M} \ne 0$, a constant $\lambda$, and functions $b_k$, $c_k$ as in
(\ref{bc}), such that $H(\tilde{M}, \tilde{X}, u_x, u_y, u_z) \geqslant H(\tilde{M}, \tilde{X}, v_x, v_y, v_z)$ for
every $v_x$, $v_y$, $v_z$ such that $v_x^2 + v_y^2 + v_z^2 \leqslant \gamma^2$. Moreover, the Pontryagin Hamiltonian
is constant and equal to $\lambda$.
\end{theorem}

Controls satisfying the conditions of Theorem \ref{the1} are called {\it extremals}, and they are candidate
optimal controls, since the PMP is a necessary condition for optimality. They are called {\it normal} extremals if
$\lambda \ne 0$, {\it abnormal} otherwise. The
arcs of the extremal trajectories where the Pontryagin Hamiltonian does not explicitly depend on the controls
are called singular, and in this specific case they are described by $b_x \equiv b_y \equiv b_z \equiv 0$. In
general an extremal trajectory can contain both singular and nonsingular arcs. We shall call {\it singular} a trajectory
which contains at least one singular arc, {\it nonsingular} otherwise.

The procedure to find the time optimal control consists of computing all the extremals, and then compare the values of
the time needed to reach the desired final condition and choose the control which gives the minimum time.

For both singular and nonsingular arcs, and, in fact, for every control, the functions defined in (\ref{bc})
satisfy the following differential equations,
\begin{eqnarray}\label{syst}
  \dot{b}_x &=& - \omega_0 c_y - u_z b_y + u_y b_z, \nonumber \\
  \dot{b}_y &=& \omega_0 c_x + u_z b_x - u_x b_z, \nonumber \\
  \dot{b}_z &=& u_x b_y - u_y b_x, \nonumber \\
  \dot{c}_x &=& - \omega_0 b_y + u_y c_z - u_z c_y, \nonumber \\
  \dot{c}_y &=& \omega_0 b_x - u_x c_z + u_z c_x, \nonumber \\
  \dot{c}_z &=& u_x c_y - u_y c_x.
\end{eqnarray}
They can be derived by differentiating (\ref{bc}), taking into account the evolution (\ref{schr3}), and the commuting
relations (\ref{comm}).

For some given initial conditions $b_k (0)$, and $c_k (0)$, $k = x, y, z$ the final conditions $b_k (t_f)$ and $c_k (t_f)$,
are obtained by taking $X_f = X(t_f)$ in (\ref{bc}). They read
\begin{equation}\label{so3}
b_k (t_f) = \sum_l \mathcal{X}_{kl} b_l (0), \qquad c_k (t_f) = \sum_l \mathcal{X}_{kl} c_l (0),
\end{equation}
where $\mathcal{X} \in SO(3)$ is a matrix representation of the element associated with $X_f$. This follows
from the fact that $SU(2)$ is the double covering group of $SO(3)$.

\noindent It turns out that (\ref{pont}) is not the only constant of motion on extremal trajectories.

\begin{proposition}\label{rem2}
System (\ref{syst}) admits the following integrals of motion~\footnote{The multiplicative constants are introduced for
further reference.}:
\begin{equation}\label{cons}
E = \frac{\omega^2_0}{2} \sum_{k = x, y, z} (b_k^2 + c_k^2), \qquad L = \omega_0 \sum_{k = x, y, z} b_k c_k.
\end{equation}
\end{proposition}
\noindent {\it Proof:} The fact that $\dot{E} = \dot{L} = 0$ is a direct consequence of (\ref{syst}). \\

In the next section we will provide a useful interpretation of these constants. To complete this section,
we derive some results which will be needed in the following. First of all, we characterize
the structure of extremal control strategies. On nonsingular arcs, since the Pontryagin Hamiltonian is linear in $v_x$, $v_y$ and $v_z$, its
maximum in the set $v_x^2 + v_y^2 + v_z^2 \leqslant \gamma^2$ must be on the border of this set.
By applying the Lagrange multipliers method we find that the extremal controls satisfy
\begin{equation}\label{optc}
    u_k = \gamma \frac{b_k}{\mu_b}, \qquad k = x, y, z
\end{equation}
where $\mu_b = \sqrt{b_x^2 + b_y^2 + b_z^2}$. On nonsingular arcs, $\mu_b$ can vanish only in isolated points.
From standard theorems on ordinary differential equations, it follows that $b_k$ must be continuous, and then
the controls $u_k$ must be piecewise continuous.

\begin{remark}\label{rem3}
At the beginning of this section, we have proved that two target operators $X_f$ and $\hat{X}_f$ satisfying
(\ref{Xtil}) are reached in the same minimum time. One possible interpretation of this result is to
think of (\ref{Xtil}) as a symmetry transformation which preserves the Lie algebraic structure of the problem.
Accordingly, if the matrix  $M \in \mathfrak{l}$ appearing in the PMP is associated with $X_f$, the
corresponding operator $\hat{M} \in \mathfrak{l}$ associated with $\hat{X}_f$, must be given by
\begin{equation}\label{Mtil}
    \hat{M} = e^{i \varphi S_z} M e^{-i \varphi S_z}, \qquad \varphi \in \mathbb{R}.
\end{equation}
The coefficients (\ref{bc}) are transformed as
$\hat{b}_x = b_x \cos{\varphi} + b_y \sin{\varphi}$, $\hat{b}_y = -b_x \sin{\varphi} + b_y \cos{\varphi}$ and
$\hat{b}_z = b_z$, where $\hat{b}_k$, $k = x, y, z$, are associated with $\hat{M}$ (the same transformation
describes the map of the $c_k$ coefficients into $\hat{c}_k$), and the Pontryagin Hamiltonian is unchanged.
\end{remark}


\section{The costate dynamics as a central-force problem}\label{sec3}

In this section we reinterpret the costate dynamics, i.e. Eq.s (\ref{syst}), as a central-force problem.
This parallel provides a useful bridge between our problem and a well know scenario in classical mechanics,
which has been deeply investigated in the past. While we will find this connection especially useful in the
special case $X_f = X_{SWAP}$ or equivalent operator, we believe that it is significant for more general final conditions. Therefore, for sake
of completeness, initially we consider a generic target operator, and then we specialize our analysis to the
case-study of $X_f = X_{SWAP}$.

Assume we are on a nonsingular extremals. By using the general form of the controls (\ref{optc}),
we see from (\ref{syst}) that $\dot{b}_z \equiv 0$, and then $b_z = b_z (0)$ is a constant. This implies that
\begin{equation}\label{mube}
    \dot{\mu}_b = \frac{1}{\mu_b} (b_x \dot{b}_x + b_y \dot{b}_y) = - \frac{\omega_0}{\gamma} \dot{c}_z,
\end{equation}
which is equivalent to the constancy of the Pontryagin Hamiltonian on nonsingular extremals:
\begin{equation}\label{pont2}
    H(\tilde{M}, \tilde{X}, u_x, u_y, u_z) =  \lambda = \omega_0 c_z + \gamma \mu_b = \omega_0 c_z (0) + \gamma \mu_b (0).
\end{equation}
This result is embodied in the last equation in (\ref{syst}). The other equations reduce to
\begin{eqnarray}\label{syst2}
  \dot{b}_x &=& - \omega_0 c_y, \nonumber \\
  \dot{b}_y &=& \omega_0 c_x, \nonumber \\
  \dot{c}_x &=& - \omega_0 b_y + \frac{\gamma}{\mu_b} \big(b_y c_z - b_z (0) c_y\big), \nonumber \\
  \dot{c}_y &=& \omega_0 b_x - \frac{\gamma}{\mu_b} \big(b_x c_z - b_z (0) c_x\big),
\end{eqnarray}
which, on account of (\ref{pont2}), can be written as
\begin{eqnarray}\label{syst3}
    \ddot{b}_x &=& - (\omega_0^2 + \gamma^2) b_x + \gamma \lambda
    \frac{b_x}{\mu_b} - \gamma b_z (0) \frac{\dot{b}_y}{\mu_b} \nonumber \\
    \ddot{b}_y &=& - (\omega_0^2 + \gamma^2) b_y + \gamma \lambda
    \frac{b_y}{\mu_b} + \gamma b_z (0) \frac{\dot{b}_x}{\mu_b}
\end{eqnarray}
These non-linear equations can be interpreted as modeling the motion of a material point of unit mass, described
by a ``position'' vector ${\bf b} = (b_x, b_y) = (b \cos{\theta}, b \sin{\theta})$, in Cartesian or polar
coordinates, respectively. This point is driven by a central force, the first and second terms in the right
hand side of (\ref{syst3}), plus a term dependent on ``velocity'' (the last contribution in both equations). By
using a compact notation,
\begin{equation}\label{cent}
	\ddot{{\bf b}} = {\bf F} ({\bf b}) + {\bf f} ({\bf b}, \dot{{\bf b}}),
\end{equation}
where
\begin{equation}\label{forc}
	{\bf F} ({\bf b}) = -(\omega_0^2 + \gamma^2) {\bf b} + \gamma \lambda
	\frac{{\bf b}}{\sqrt{b^2 + b_z^2 (0)}}, \quad {\bf f} ({\bf b}, \dot{{\bf b}}) =
    -i \gamma b_z (0) \sigma_y \frac{\dot{{\bf b}}}{\sqrt{b^2 + b_z^2 (0)}}.
\end{equation}
The velocity-dependent term is not a dissipative contribution. In fact, it is possible to interpret the integral
of motion $E$ of the system, given in Proposition~\ref{rem2}, as the associated {\it energy}. This can be seen by
introducing the potential associated to the radial part of the force,
\begin{equation}\label{pot}
	U (b) = \frac{1}{2} (\omega_0^2 + \gamma^2) \left( b^2 + b_z^2 (0) \right) -
    \gamma \lambda \sqrt{b^2 + b_z^2 (0)} +  \frac{1}{2} \lambda^2,
\end{equation}
with ${\bf F} ({\bf b}) = -{\rm grad}_{\bf b} U(b)$. It follows that
\begin{equation}\label{cons2}
	E = U (b) + \frac{1}{2} \Big( \dot{b}^2 + (b \dot{\theta})^2 \Big),
\end{equation}
which justifies the interpretation of $E$ as the energy (potential plus kinetic energy). The energy is conserved,
and there is no dissipation.

Since we have found $b_z \equiv b_z (0)$, and by virtue of (\ref{so3}), we can conclude that $b(t_f) = b(0)$.
Since $E$ is an integral of motion, we must also have $c_z (t_f) = c_z (0)$ (alternatively, this is a
consequence of the invariance of the Pontryagin Hamiltonian).

\subsection{The case $X_f = X_{SWAP}$}

If we assume $X_f = X_{SWAP}$ or any equivalent operator parameterized by $\varphi$ according to
Proposition~\ref{rem1}, we have some simplifications. Therefore, here and in the following
sections we will limit our attention to this special case, and then comment about the general case again in Section~\ref{sec7}.
Since we consider special unitary target operators, we conventionally write $X_{SWAP} = i \sigma_y$, and then
the aforementioned family of equivalent operators is given by
\begin{equation}\label{swap}
    X_f = \left(
              \begin{array}{cc}
                0 & e^{i \varphi} \\
                -e^{-i \varphi} & 0 \\
              \end{array}
            \right),
\end{equation}
where $\varphi$ is an arbitrary real number, which can be taken in the interval $[0, 2 \pi)$
without loss of generality. The $SO(3)$ matrix introduced in (\ref{so3}) is given by
\begin{equation}\label{so3bis}
\mathcal{X} = \left(
                \begin{array}{ccc}
                  -\cos{2\varphi} & \sin{2\varphi} & 0 \\
                  \sin{2\varphi} & \cos{2\varphi} & 0 \\
                  0 & 0 & -1 \\
                \end{array}
              \right),
\end{equation}
which is a rotation of angle $\pi$ about the axis defined by the unit vector $(\sin{\varphi}, \cos{\varphi}, 0)$.
For the final coordinates we must have
\begin{eqnarray}\label{infin}
    b_x (t_f) &=& -  b_x (0) \cos{2\varphi} + b_y (0) \sin{2\varphi}, \nonumber \\
    b_y (t_f) &=& b_x (0) \sin{2\varphi} + b_y (0) \cos{2\varphi}, \nonumber \\
    b_z (t_f) &=& -b_z (0),
\end{eqnarray}
or, in polar coordinates, $b (t_f) = b (0)$ (as we already know), and $\theta(t_f) = \pi - \theta(0) - 2 \varphi$.
Similar relations hold for $c_k (t_f)$. Consistency with the former constraints requires $b_z (t_f) = b_z (0) = 0$
and $c_z (t_f) = c_z (0) = 0$. This implies that the Pontryagin Hamiltonian equals $\lambda = \gamma b (0)$, the
velocity-dependent term disappears, ${\bf f}({\bf b}, \dot{{\bf b}}) = {\bf 0}$, and the central force simplifies to
\begin{equation}\label{forc2}
	{\bf F} ({\bf b}) = -(\omega_0^2 + \gamma^2) {\bf b} + \gamma \lambda
	\frac{{\bf b}}{b}.
\end{equation}
The associated potential is given by
\begin{equation}\label{pot2}
	U (b) = \frac{1}{2} (\omega_0^2 + \gamma^2) b^2 - \gamma \lambda b,
\end{equation}
which belongs to the class of integrable potentials, with solutions given in terms of elliptic integrals. Initial and final
positions and velocities are related by an orthogonal transformation, the upper $2 \times 2$ diagonal block in (\ref{so3bis}).
Moreover, the integral of motion $L$, given in Proposition~\ref{rem2},  takes the form
\begin{equation}\label{cons3}
	L = b^2 \dot{\theta},
\end{equation}
and then it can be interpreted as the {\it angular momentum} of the system.

Because of its particular simplicity, which highly limits the need of numerical analysis, in the reminder of this paper
we will fully explore this case. As anticipated in the Introduction, we now separately consider nonsingular and singular extremals.


\section{Nonsingular extremals for $\rm X_f = X_{SWAP}$}\label{sec4}

In this section we limit our attention to nonsingular extremals, for which $\mu_b$ can vanish
only at isolated points. Following the analysis of the previous section, we know that it is possible
to express the costate dynamics in terms of elliptic integrals. Unfortunately, this is not enough
to provide useful expressions for the integrals of (\ref{schr2}), that is, suitable expressions
for the comparison of the extremal trajectories. Nonetheless, numerical computations suggests that the minimum
time trajectories are characterized by $L = 0$. We are not able to prove this result analytically, but we have
found numerical evidence of its validity. We refer to Appendix~\ref{app1} for more details on the numerical
analysis supporting this fact.
\begin{conjecture}
If $X_f = X_{SWAP}$, and an extremal control strategy is optimal, then the corresponding trajectory satisfies $L = 0$.
\end{conjecture}
Under this assumption, the dynamics of the costate greatly simplifies, and it is possible to compute explicit expressions for
the evolution in the Lie group.

The following argument is based on Proposition~\ref{rem3}. If we assume that the optimal control strategy leading to $X_f = X_{SWAP}$
requires ${\mathbf b} (0) = (b_x (0), b_y (0))$, then an equivalent operator, differing from $X_f$ by a phase $\varphi$ in the off-diagonal
elements, is associated with the counter-clockwise rotation by an  angle $\varphi$ of ${\mathbf b} (0)$. In particular, one of
these operators has to be associated with ${\mathbf b} (0) = (b_x (0), 0)$, with $b_x (0) \geqslant 0$.
According to Proposition~\ref{rem1}, there is no loss of generality in choosing this operator as $X_f$:
the general analysis of any operator equivalent to SWAP will follow~\footnote{Alternatively, we could study the costate
dynamics by using the polar coordinates $b$ and $\theta$. While this approach makes more transparent the symmetry of the problem
(which is, basically, independence from $\theta$), we prefer to present our analysis in terms of $b_x$ and $b_y$, because
they have a smooth evolution. This is not the case for $b$ (whose derivative changes sign when $b = 0$) and
$\theta$ (which flips between two fixed values).}.

Since we have assumed $L = 0$, we can write ${\mathbf b}(t) = (b_x(t), 0)$, and then
$b_y (t) \equiv 0$, which implies $b_y (t_f) = 0$. But validity of (\ref{infin}) imposes that $\sin{2 \varphi} = 0$, which can be
satisfied either by $\varphi = 0$ or $\varphi = \frac{\pi}{2}$. These two cases correspond to $X_f = i \sigma_y$ and
$X_f = i \sigma_x$, respectively. Before analyzing them, we determine the structure of
extremal control strategies for nonsingular arcs.

Since $\mu_b = \vert b_x \vert$, we can rewrite the first equation in (\ref{syst3}) as
\begin{equation}\label{evbx}
    \ddot{b}_x = - (\omega_0^2 + \gamma^2) b_x + {\rm sign} (b_x) \gamma^2 b_x (0),
\end{equation}
with general solution
\begin{equation}\label{evbxsol}
    b_x (t) = A_j \cos{\omega t} + B_j \sin{\omega t} + {\rm sign} (b_x) \left(\frac{\gamma}{\omega}\right)^2 b_x (0),
\end{equation}
where $\omega^2 = \omega_0^2 + \gamma^2$. The coefficients $A_j$ and $B_j$, $j = 0, 1, 2, \ldots$
are determined by the initial conditions, and
by the requirement that $b_x (t)$ and $\dot{b}_x (t)$ are continuous functions (whenever $b_x$ vanishes, new coefficients must
be taken into account). The initial pair of coefficients is given by
\begin{equation}\label{ABco}
    A_0 = \left(\frac{\omega_0}{\omega}\right)^2 b_x (0), \qquad B_0 = \pm \frac{\omega_0}{\omega} \sqrt{1 - b_x (0)^2},
\end{equation}
and $B_0$ has been written under the assumption that $E = \frac{\omega^2_0}{2}$, since
the value of this constant does not affect the optimal control problem.
The sign of $B_0$ is the same as the sign of $\dot{b}_x (0)$. We have
two possible solutions (\ref{evbxsol}), depending on the sign of $B_0$. All the other coefficients $A_j$
and $B_j$ are completely determined by the continuity requirements, and their explicit expressions
are not needed for further developments. Following (\ref{optc}), the optimal control strategy is necessarily bang-bang, with
\begin{equation}\label{optc2}
    u_x = \gamma \, {\rm sign} (b_x), \qquad u_y = u_z = 0.
\end{equation}
If they exist, the switching times are given by the zeros of (\ref{evbxsol}), which depend on $b_x (0)$.
We denote these switching times by $\tilde{t}_k = \tilde{t} + k \bar{t}$, $k = 0, 1, 2, \ldots$, with $\bar{t} \geqslant \tilde{t} \geqslant 0$.
The zeros of (\ref{evbxsol}) satisfy
\begin{eqnarray}\label{ttil}
    \cos{\omega t_{\pm}} &=& \frac{- A_0 C_0 \pm B_0 \sqrt{A_0^2 + B_0^2 - C_0^2}}{A_0^2 + B_0^2}, \nonumber \\
    \sin{\omega t_{\pm}} &=& \frac{- B_0 C_0 \mp A_0 \sqrt{A_0^2 + B_0^2 - C_0^2}}{A_0^2 + B_0^2},
\end{eqnarray}
with $A_0$ and $B_0$ as in (\ref{ABco}), and $C_0 = \left(\frac{\gamma}{\omega}\right)^2 b_x (0)$. We can assume that
$t_+$ and $t_-$ are, in absolute value, the smaller and next to smaller zeros, not necessarily in this order. If $b_x (0) \ne 0$,
by considering both positive or negative values for $B_0$ in (\ref{ttil}), we find that
$\sin{\omega t_+} \leqslant \sin{\omega t_-}$. Moreover, for $B_0 \geqslant 0$ we find $\cos{\omega t_+} \geqslant \cos{\omega t_-}$,
$\cos{\omega t_-} \leqslant 0$ and $\sin{\omega t_+} \leqslant 0$, and for $B_0 < 0$, we find $\cos{\omega t_+} \leqslant \cos{\omega t_-}$,
$\cos{\omega t_+} \leqslant 0$ and $\sin{\omega t_-} \geqslant 0$. From these properties we can conclude that, in any case,
$t_- \geqslant 0$ and $t_+ \leqslant 0$, so that
\begin{equation}\label{ttild}
	\cos{\omega \tilde{t}} = \cos{\omega t_-}, \qquad \sin{\omega \tilde{t}} = \sin{\omega t_-}
\end{equation}
and $\bar{t} = (t_- - t_+)$, leading to
\begin{equation}\label{tbar}
    \cos{\omega \bar{t}} = \frac{2 C_0^2 - A_0^2 - B_0^2}{A_0^2 + B_0^2}, \qquad
    \sin{\omega \bar{t}} = - \frac{2 C_0 \sqrt{A_0^2 + B_0^2 - C_0^2}}{A_0^2 + B_0^2}.
\end{equation}

We don't need explicit expressions of $\tilde{t}$ and $\bar{t}$ in terms of $\omega_0$, $\gamma$ and $b_x (0)$, and Eq.s (\ref{tbar}),
(\ref{ttild}) and (\ref{ttil}) will be sufficient to fully specify the control strategies in the case of interest for this work. In the special case
$b_x (0) = 0$, we have $A_0 = C_0 = 0$ and $B_0 = \pm \frac{\omega_0}{\omega}$,
from which we can evaluate $\tilde{t} = \bar{t} = \frac{\pi}{\omega}$.
For further reference, we observe the following:
\begin{proposition}\label{remadd1}
From $\sin{\omega \bar{t}} \leqslant 0$ we have that $\frac{\pi}{\omega} \leqslant \bar{t} \leqslant \frac{2 \pi}{\omega}$. Moreover, $0 \leqslant \tilde{t} \leqslant \frac{\pi}{\omega}$. \\
\end{proposition}

\noindent In the following, we separately consider the two aforementioned cases $X_f = i \sigma_y$ or $X_f = i \sigma_x$.

\subsection{Case with $X_f = i \sigma_y$}

By using $\varphi = 0$ in (\ref{infin}), we derive the final condition $b_x (t_f) = - b_x (0)$. An explicit
computation proves that, if $b_x (0) = 0$ and the control $u_x$ is constant, it is impossible to reproduce $X_f$.
Therefore, control strategies associated with these nonsingular extremals have an odd number of switching
times. We adopt the following definitions:
\begin{equation}\label{Spm}
    S_+ = \frac{1}{\omega} (\gamma S_x + \omega_0 S_z), \qquad
    S_- = \frac{1}{\omega} (\gamma S_x - \omega_0 S_z),
\end{equation}
and with  $\omega = \sqrt{\omega_0^2 + \gamma^2}$. By integrating the dynamics (\ref{schr2}), we find
\begin{eqnarray}\label{solX}
    X_1 (t) &=& Y^{\dagger}_- (\tilde{t}) \Big( Y_+ (\bar{t}) Y^{\dagger}_- (\bar{t}) \Big)^n Y_+ (\tilde{t}), \nonumber \\
    X_2 (t) &=& Y^{\dagger}_+ (\tilde{t}) \Big( Y_- (\bar{t}) Y^{\dagger}_+ (\bar{t}) \Big)^n Y_- (\tilde{t}),
\end{eqnarray}
where $n = 0, 1, 2, \ldots$, and we have used the notation
\begin{equation}\label{Y+-}
    Y_{+} (t) = e^{- i \omega t S_{+}}, \qquad Y_{-} (t) = e^{- i \omega t S_{-}}.
\end{equation}
By requiring that $X_1 (t_f) = - X_2 (t_f) = \pm X_f$, from the pair of equations (\ref{solX}), we obtain
\begin{equation}\label{Y+Y-}
    \Big( Y_+ (\bar{t}) Y^{\dagger}_- (\bar{t}) \Big)^n = \pm Y_- (\tilde{t}) X_f Y^{\dagger}_+ (\tilde{t}),
\end{equation}
and the total time of evolution on these nonsingular trajectories is $t_f = 2 (\tilde{t} + n \bar{t})$.
It turns out that the other possibility, $X_1 (t_f) = X_2 (t_f) = \pm X_f$, is inconsistent. To compactly
present the treatment of this case, we find convenient to define two functions $\alpha = \alpha (\omega_0, \gamma, \bar{t})$
and $\beta = \beta (\omega_0, \gamma, \bar{t})$ satisfying
\begin{eqnarray}\label{albe}
    \sin{\alpha} = \left( \frac{\omega_0}{\omega} \right) \sin{\omega \bar{\tau}}, &\quad&
    \cos{\alpha} = \sqrt{1 - \left( \frac{\omega_0}{\omega} \right)^2 \sin^2{\omega \bar{\tau}}}, \nonumber \\
    \sin{\beta} = - \sec{\alpha} \cos{\omega \bar{\tau}}, &\quad& \cos{\beta} = \frac{\gamma}{\omega}
    \sec{\alpha} \sin{\omega \bar{\tau}},
\end{eqnarray}
where $\bar{\tau} = \frac{\bar{t}}{2}$. These definitions are meaningful only when $\cos{\alpha} \ne 0$, which is always
the case if $\gamma \ne 0$.  To prove that they are consistent is a lengthy but
standard problem of trigonometry. We can now write in compact form
\begin{equation}\label{Y+Y-bis}
    Y_+ (\bar{t}) Y^{\dagger}_- (\bar{t}) = \cos{2\alpha} I + i \sin{2\alpha} \, (\cos{\beta} \sigma_y
    + \sin{\beta} \sigma_z),
\end{equation}
from which it follows that we can express the left-hand side of (\ref{Y+Y-}) as
\begin{equation}\label{lhs}
    \Big( Y_+ (\bar{t}) Y^{\dagger}_- (\bar{t}) \Big)^n = \cos{2 n \alpha} I + i \sin{2 n \alpha} \, (\cos{\beta} \sigma_y
    + \sin{\beta} \sigma_z).
\end{equation}
On the other side, the right-hand side of (\ref{Y+Y-}) is computed by considering (\ref{Y+-}) and the fact that $X_f = i \sigma_y$:
\begin{equation}\label{rhs}
    Y_- (\tilde{t}) X_f Y^{\dagger}_+ (\tilde{t}) = \mp 2 \frac{\omega_0 \gamma}{\omega^2} \sin^2{\omega \tilde{\tau}} I
    \pm i \left( 1 - 2 \left( \frac{\gamma}{\omega} \right)^2 \sin^2{\omega \tilde{\tau}} \right) \sigma_y \pm i \frac{\gamma}{\omega}
    \sin{2 \omega \tilde{\tau}} \sigma_z.
\end{equation}
Now, by comparing (\ref{lhs}) and (\ref{rhs}), we obtain the following set of equations:
\begin{equation}\label{system}
    \left\{ \begin{array}{l}
              \cos{2 n \alpha} = \mp 2 \frac{\omega_0 \gamma}{\omega^2} \sin^2{\omega \tilde{\tau}} \\
              \sin{2 n \alpha} \cos{\beta} = \pm \left( 1 - 2 \left( \frac{\gamma}{\omega} \right)^2 \sin^2{\omega \tilde{\tau}} \right) \\
              \sin{2 n \alpha} \sin{\beta} = \pm \frac{\gamma}{\omega} \sin{2 \omega \tilde{\tau}}
            \end{array}
    \right.
\end{equation}
which can be analytically solved in simple cases. The details of the computation when $n = 0, 1, 2$ are reported in Appendix~\ref{app2},
as well as the values of $\tilde{t}$ and $\bar{t}$, which fully characterize the control strategy, the corresponding final time $t_f$,
and the range of existence of these extremals in terms of $\omega_0$ and $\gamma$.

\subsection{Case with $X_f = i \sigma_x$}

In this case $\varphi = \frac{\pi}{2}$, and, from (\ref{infin}), the final condition is $b_x (t_f) = b_x (0)$. It is possible
to prove that a constant control strategy is possible only if $\omega_0 = 0$, which has been excluded at the beginning.
Therefore, any control strategy for this family of nonsingular extremals must have at least one switching time, and
in fact, there must be an even number of them. The solutions of (\ref{schr2}) read
\begin{eqnarray}\label{solX2}
    X_1 (t) &=& Y_+ (\tilde{t}) Y^{\dagger}_- (\bar{t}) \Big( Y_+ (\bar{t}) Y^{\dagger}_- (\bar{t}) \Big)^n Y_+ (\tilde{t}), \nonumber \\
    X_2 (t) &=& Y_- (\tilde{t}) Y^{\dagger}_+ (\bar{t}) \Big( Y_- (\bar{t}) Y^{\dagger}_+ (\bar{t}) \Big)^n Y_- (\tilde{t}),
\end{eqnarray}
where $n = 0, 1, 2, \ldots$. Working as before, and considering the cases $X_1 (t_f) = X_f$
and $X_2 (t_f) = \pm X_f$, we find the necessary conditions
\begin{eqnarray}\label{Y+Y-2}
    Y^{\dagger}_- (\bar{t}) \Big( Y_+ (\bar{t}) Y^{\dagger}_- (\bar{t}) \Big)^n &=& Y^{\dagger}_+ (\tilde{t}) X_f Y^{\dagger}_+
    (\tilde{t}), \nonumber \\
    \Big( Y_+ (\bar{t}) Y^{\dagger}_- (\bar{t}) \Big)^n Y_+ (\bar{t})  &=& \mp Y_- (\tilde{t}) X_f Y_- (\tilde{t}),
\end{eqnarray}
and the total time for the transition is given by $t_f = 2 \tilde{t} + (2n + 1) \bar {t}$.
By using (\ref{lhs}) and (\ref{Y+-}) we compute
\begin{eqnarray}\label{lhs2}
  Y^{\dagger}_- (\bar{t}) \Big( Y_+ (\bar{t}) Y^{\dagger}_- (\bar{t}) \Big)^n &=& - \cos{(2n + 1)\alpha} (\sin{\beta} I - i \cos{\beta} \sigma_x)
  - i \sin{(2n + 1)\alpha} \sigma_z, \nonumber \\
  \Big( Y_+ (\bar{t}) Y^{\dagger}_- (\bar{t}) \Big)^n Y_+ (\bar{t}) &=& - \cos{(2n + 1)\alpha} (\sin{\beta} I + i \cos{\beta} \sigma_x)
  - i \sin{(2n + 1)\alpha} \sigma_z,
\end{eqnarray}
and the right hand sides of (\ref{Y+Y-2}) can be written as
\begin{eqnarray}\label{rhs2}
    Y^{\dagger}_+ (\tilde{t}) X_f Y^{\dagger}_+ (\tilde{t}) &=& - \frac{\gamma}{\omega} \sin{2 \omega \tilde{\tau}} I
    + i \left(1 - 2 \left( \frac{\gamma}{\omega} \right)^2 \sin^2{\omega \tilde{\tau}}\right) \sigma_x - 2 i
    \frac{\omega_0 \gamma}{\omega^2} \sin^2{\omega \tilde{\tau}} \sigma_z, \nonumber \\
    Y_- (\tilde{t}) X_f Y_- (\tilde{t}) &=& \frac{\gamma}{\omega} \sin{2 \omega \tilde{\tau}} I
    + i \left(1 - 2 \left( \frac{\gamma}{\omega} \right)^2 \sin^2{\omega \tilde{\tau}}\right) \sigma_x + 2 i
    \frac{\omega_0 \gamma}{\omega^2} \sin^2{\omega \tilde{\tau}} \sigma_z,
\end{eqnarray}
where $\tilde{\tau} = \frac{\tilde{t}}{2}$.
By imposing equality of (\ref{lhs2}) and (\ref{rhs2}), we find that $X_1 (t_f) = - X_2 (t_f)$ is inconsistent, and
$X_1 (t_f) = X_2 (t_f) = \pm X_f$ is satisfied under the constraints
\begin{equation}\label{system2}
    \left\{ \begin{array}{l}
              \cos{(2 n + 1) \alpha} \sin{\beta} = \pm \frac{\gamma}{\omega} \sin{2 \omega \tilde{\tau}} \\
              \cos{(2 n + 1) \alpha} \cos{\beta} = \pm \left( 1 - 2 \left( \frac{\gamma}{\omega} \right)^2
              \sin^2{\omega \tilde{\tau}} \right) \\
              \sin{(2 n + 1) \alpha} = \pm 2 \frac{\omega_0 \gamma}{\omega^2}  \sin^2{\omega \tilde{\tau}}
            \end{array}
    \right.
\end{equation}
In general, the solution of this system can be performed numerically. Nonetheless,
in the simple cases $ n = 0, 1$ analytical solutions are possible. We refer to Appendix~\ref{app3}
for the details.


\section{Singular extremals for $\rm X_f = X_{SWAP}$}\label{sec5}

In this section we consider the case of singular extremals, that is, extremal trajectories containing at least one singular arc.
Although we limit our attention to $X_f = X_{SWAP}$, the analysis of singular extremals can be performed for any final target operator.
For sake of simplicity, this generalization will be presented elsewhere.

In Section~\ref{sec2} we have seen that, in the present context, the condition for a singular arc is $b_x \equiv b_y \equiv b_z \equiv 0$.
On this arc, $\dot{b}_x \equiv \dot{b}_y \equiv \dot{b}_z \equiv 0$, which implies from (\ref{syst}) that $c_x \equiv c_y \equiv 0$, and
then $\dot{c}_z \equiv 0$, that is, $c_z$ is a non zero constant. It cannot vanish because, in the Pontryagin Maximum principle,
$\tilde{M} \ne 0$.  Moreover, to satisfy (\ref{syst}), it must be $u_x \equiv u_y \equiv 0$, and the only control possibly different from zero is $u_z$.

On these extremal trajectories it must be $L = 0$, since this condition holds on any embedded singular arc, and $L$ is constant
everywhere. On singular arcs, the Pontryagin Hamiltonian gives $\lambda = \omega_0 c_z$, therefore the
expression (\ref{pont2}) is generically valid. We maintain our simplifying assumption $E = \frac{\omega^2_0}{2}$ (which is true without loss of generality).

Now we prove that any singular arc must be preceded and followed by nonsingular arcs.
\begin{proposition}\label{rem4}
If $X_f$ is the SWAP operator (or an equivalent operator,
according to Proposition~\ref{rem1}), a singular arc cannot be at the beginning, or at the end, of an extremal trajectory.
\end{proposition}
\noindent {\it Proof:} Assume that a singular arc is at the beginning of the trajectory. The expression of ${\mathcal{X}}$ associated to this class of final
operators is given in (\ref{so3bis}). It follows that $c_z(t_f) = - c_z(0)$. Since we have chosen $E = \frac{\omega^2_0}{2}$, it
must be $c_z (0) = 1$,  and then $c_z (t_f) = -1$, which in turn implies $\mu_b (t_f) = 0$. But this is inconsistent with the constancy
of the Pontryagin Hamiltonian, since its values at $t = 0$ and $t = t_f$ are different. A completely analogous argument rules out
extremal trajectories ending with singular arcs. \\

As we did before, by relying on Proposition~\ref{rem3} we can assume without loss of generality that ${\mathbf b}(0) = (b_x (0), 0)$,
and $b_x (0) \geqslant 0$. The analysis of the dynamics of ${\mathbf b}(t)$ on non-singular arcs when $L = 0$ has been detailed in the previous section. We
can adapt this treatment to the present case, by considering the additional requirement that singular arcs should be smoothly
connected to nonsingular arcs. In the $(b_x, b_y)$ plane, nonsingular arcs follow segments through the origin, and singular arcs
correspond to a point at rest in the origin. Therefore, to connect to a singular arc, the preceding arc must end with
$b_x (t) = 0$ and $\dot{b}_x (t) = 0$. For the first singular arc, $b_x (t) = 0$ gives (\ref{ttil}), and $\dot{b}_x (0) = 0$ means
\begin{equation}\label{ttil2}
	- \omega A_0 \sin{\omega t_{\pm}} + \omega B_0 \cos{\omega t_{\pm}} = 0.
\end{equation}
Joint consideration of these constraints leads to $b_x (0) = \frac{\omega_0}{\gamma}$: this is the only initial condition such that
the system admits singular arcs. Since $b_x (0) \leqslant 1$, we conclude that it is possible to have extremals with singular
arcs only if $\gamma \geqslant \omega_0$, which physically corresponds to a control power of the same or more magnitude than the interaction. Correspondingly, we have
\begin{equation}\label{incondo}
	A_0 = \frac{\omega_0^3}{\gamma \omega^2}, \qquad B_0 = \pm \frac{\omega_0}{\gamma \omega} \sqrt{\gamma^2 - \omega_0^2},
	\qquad C_0 = \frac{\gamma \omega_0}{\omega^2},
\end{equation}
and both the smaller and next to smaller zeros of $b_x (t)$ satisfy
\begin{equation}\label{cosi}
	\cos{\omega t_\pm} = - \frac{A_0}{C_0}, \qquad \sin{\omega t_\pm} =  - \frac{B_0}{C_0}.
\end{equation}
Without loss of generality we assume that $t_+ > 0$, therefore $\omega (t_+ - t_-) = 2  \pi$ and the relevant times are  $\tilde{t} = t_+$,
$\bar{t} = \frac{2 \pi}{\omega}$. We can now prove that multiple singular arcs are not possible  if the  extremal trajectory trajectory is optimal.
\begin{proposition}\label{rem5}
A time  optimal trajectory may  contain at most one singular arc.
\end{proposition}
\noindent {\it Proof:} If a non-singular arc is in between two singular arcs, the corresponding evolution requires a time $k \tau$, for
some positive integer $k$. The evolution operators are given by $Y_+ ( k \bar{t}) = - I$ and $Y^{\dagger}_- ( k \bar{t}) = - I$, a
ccording to (\ref{Y+-}). These contributions are inconsistent with the requirement of optimality. They amount to no evolution in positive time. \\

Propositions~\ref{rem4} and \ref{rem5} significantly constrain the form of any possible singular optimal candidates: it must be the succession
of evolutions on a nonsingular arc for time $\tilde{t}$, on a singular arc for time $t^{\prime}$, and finally on a nonsingular arc, again for
time $\tilde{t}$, because of the requirements on the final conditions $b_k (t_f)$ and $c_k (t_f)$. In the $(b_x, b_y)$ plane, the state starts
from the point $(\frac{\omega_0}{\gamma})$ and move to $(0,0)$ along a straight line, during the first nonsingular evolution; then it remains in
$(0,0)$ during the singular evolution; finally, it moves to $(- \frac{\omega_0}{\gamma} \cos{2 \varphi}, \frac{\omega_0}{\gamma} \sin{2 \varphi})$
along a straight line, during the second nonsingular evolution. The complete evolution on the Lie group reads
\begin{equation}\label{singu}
	X_1 (t) =  Z_+ (\tilde{t}) W_+ (t^{\prime}) Y_+ (\tilde{t}), \qquad X_2 (t) =  Z_- (\tilde{t}) W_- (t^{\prime}) Y_- (\tilde{t}),
\end{equation}
with
\begin{eqnarray}\label{WZ}
	W_+ (t) = e^{-i (u_z + \omega_0)t S_z}, &\qquad& W_- (t) = e^{-i (u_z - \omega_0)t S_z}, \nonumber \\
	Z_+ (t) = e^{-i \omega t R_+}, &\qquad& Z_- (t) = e^{-i \omega t R_-},
\end{eqnarray}
where $u_z$ can be assumed to be constant (see below), and
\begin{equation}\label{Rpm}
    R_+ = \frac{1}{\omega} (u_x S_x + u_y S_y + \omega_0 S_z), \qquad
    R_- = \frac{1}{\omega} (u_x S_x + u_y S_y - \omega_0 S_z).
\end{equation}
$R_+$ and $R_-$ are functions of the constant controls on the last nonsingular arc, $u_x = - \gamma \cos{2 \varphi}$ and $u_y = \gamma \sin{2 \varphi}$.

The evaluation of the two terms in (\ref{singu}) is cumbersome and the details are reported in Appendix~\ref{app4}.
Several extremals are possible, generating any target operator equivalent to SWAP. Nonetheless, for the sake of minimizing the transfer
time, we can limit our attention to the transition to $X_f = i \sigma_y$, and compute the corresponding control strategy and transfer time
$t_f$ in this case.


\section{Optimal solutions for $\rm X_f = X_{SWAP}$}\label{sec6}

In this section we sum up the previous results and derive the time optimal control to reach the SWAP operator for the system. We have found that there is no loss of generality
considering  $b_y \equiv 0$, in which case the only possible target operations are $X_f =
i \sigma_y$ or $X_f = i \sigma_x$, and we have fully characterized the nonsingular and singular extremals,
with integral of motion $L = 0$.

According to Proposition~\ref{rem1}, to these extremals there correspond
extremals for any target operator $X_f$ equivalent to the SWAP operator. Therefore, to decide which is the
optimal solution for some given control strength and drift, it is sufficient to compare the final times associated
to these extremals, and find their minimum. We will denote the result by $t_{opt}$.
Moreover, the cases with $X_f = i \sigma_y$ and $X_f = i \sigma_x$ correspond to situations in which
there are an odd, respectively even, number of switching times for the candidate optimal control strategies.
In the following, we denote the number of switches by $s$, and use it to characterize an extremal trajectory whenever
we want to make reference to an arbitrary operator equivalent to SWAP.

First of all we derive lower and upper bounds on $t_{opt}$ on nonsingular extremals.
\begin{theorem}\label{rem6}
On nonsingular extremals, if a control strategy requires $s \geqslant 1$ switching times, then the optimal time for the
transition to $X_f = X_{SWAP}$ (or equivalent operator) must satisfy
\begin{equation}\label{bound}
(s - 1) \frac{\pi}{\omega} \leqslant t_{opt} \leqslant 2s \frac{\pi}{\omega}.
\end{equation}
\end{theorem}
\noindent {\it Proof:} This result is a direct consequence of Proposition~\ref{remadd1}. For sake of clarity, we report
the basic facts. If $X_f = i \sigma_y$, since $t_f = 2 \tilde{t}
+ 2n \bar{t}$ with $n = 0, 1,\ldots$, it must be $2n \frac{\pi}{\omega} \leqslant t_f \leqslant 2 (2n + 1) \frac{\pi}{\omega}$.
Since $b_x (t_f) = - b_x (0)$, there is an odd number of switching times, that is $s = 2n + 1$, and (\ref{bound})
follows. If $X_f = i \sigma_x$, from $t_f = 2 \tilde{t} + (2n + 1) \bar{t}$ we find $(2n + 1) \frac{\pi}{\omega} \leqslant
t_f \leqslant 4 (n + 1) \frac{\pi}{\omega}$. Since $b_x (t_f) = b_x (0)$, in this case there is an even number of switching
times, $s = 2(n + 1)$, and (\ref{bound}) is still valid. \\

We have found that there are not extremal trajectories without switching times, $s = 0$. We have analytically derived
the extremals (and corresponding control strategies) when $s \leqslant 5$. We limit our analysis to these cases, and,
in view of Theorem~\ref{rem6}, we can find the optimal solution if $t_f \leqslant \frac{5 \pi}{\omega}$,
which is the lower bound for $t_{opt}$ if $s \geqslant 6$. In fact, under this condition, the extremal trajectory that
we have not analyzed cannot outperform those which we have taken into account.

\begin{figure}[t]
  \includegraphics[width=14cm]{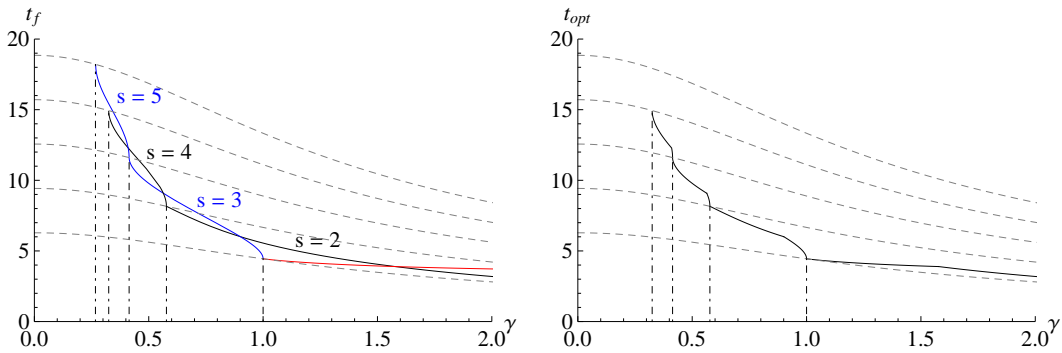}\\
  \caption{Left plot: comparison of the transition times $t_f$ for the extremal trajectories derived in this
  work, as a function of $\gamma$, when $\omega_0 = 1$. The blue and black lines are associated with the cases of even, respectively odd number of switching
  times in nonsingular extremal strategies, and the red line corresponds to the
  singular extremal. The dashed lines represents some upper and lower bounds on $t_{opt}$ as derived
  in Proposition~\ref{rem6}. More precisely, they give $k \frac{\pi}{\omega}$ for $k = 2, 3, 4, 5$ and $6$. Right plot:
  resulting optimal time in the range of values of $\gamma$ where analytical expressions for the extremals are possible.}\label{fig1}
\end{figure}
It is not possible to give the form of the time optimal control and minimum time in a  compact way since the optimal solution depends in an intricate way on the relative values of $\omega_0$
and $\gamma$. In different situations the optimal solution is given by different types of extremals, singular or nonsingular. This is illustrated in Fig.~\ref{fig1}, where the choice $\omega_0 = 1$ has been made, and the extremals which are clearly sub-optimal have been neglected.
We summarize our findings in the following theorem, where, on account of Proposition~\ref{rem1}, we provide the result for any operator
equivalent to SWAP.
\begin{theorem}
Consider the system (\ref{schr2}). Given a target operator of the form
\begin{equation}\label{eqsw}
    X_f = \left(
            \begin{array}{cc}
              0 & e^{i \varphi} \\
              -e^{- i \varphi} & 0 \\
            \end{array}
          \right)
\end{equation}
for some $\varphi \in \mathbb{R}$, the optimal control strategy steering $X_1$ and $X_2$ to $\pm X_f$
can be nonsingular or singular, depending on the relative values of $\omega_0$ and $\gamma$.

Nonsingular strategies are bang-bang, with $s$ switching times. They are given by
\begin{equation}\label{optcontrol}
    u_x = \pm \gamma \cos{\varphi}, \quad
    u_y = \mp \gamma \sin{\varphi}, \quad
    u_z = 0,
\end{equation}
and both $u_x$, $u_y$ change sign at $t_k = \tilde{t} + k \bar{t}$, with $k = 0, 1, \ldots, s$. The number $s$ depends
on the relative values of $\omega_0$ and $\gamma$. The optimal transition time is given by $t_{opt} = 2 \tilde{t} + (s - 1) \bar{t}$
for a given $s$. Analytical expressions of $\tilde{t}$ and $\bar{t}$ are provided in Appendices~\ref{app2} and \ref{app3}.

The optimal singular extremal consists of the alternation of a nonsingular arc for time $\tilde{t}$, a singular arc
for time $t^{\prime}$, and finally a nonsingular arc for time $\tilde{t}$. The optimal transition time is given by $t_{opt} =
2 \tilde{t} + t^{\prime}$, and analytical expressions of $\tilde{t}$ and $t^{\prime}$ can be found in Appendix~\ref{app4}.
The associated control strategy is given by (\ref{optcontrol}) on nonsingular arcs, and $u_x \equiv u_y \equiv u_z \equiv 0$
on the singular arc.
\end{theorem}


\section{Discussion and conclusions}\label{sec7}

The problem of controlling in minimum time a quantum bit interacting with an additional two level system can be mapped
to an equivalent problem of simultaneous control of two quantum bits under the same control field.
In this paper, we have solved this problem for the class of SWAP operators, although some of our results apply
to the case of an arbitrary final target operations as well. The application of the necessary conditions of optimality,
given by the Pontryagin Maximum Principle, leads to the consideration of two type of candidates optimal controls,
singular and nonsingular. Our analysis shows that singular extremals can be optimal only in a specific range of control
strength, and they must have a particular structure, with a singular arc between two nonsingular arcs. The dynamical
analysis benefits from known results from the central force problem. We have found that the optimal control strategy
depends in a complicated way on the relative strength of the control and the interaction. These results can be used to
synthesize the SWAP operation in minimum time in quantum computation implementations in cases where the target system is
interacting with an analogous system, and-or to synthesize the SWAP operation on two systems simultaneously.

Because of the homomorphism between $SU (2)$ and $SO (3)$, we can adapt our results to the derivation of
optimal control strategies for rotating a rigid body in $\mathbb{R}^3$ in minimum time. The class of SWAP
operators in $SU(2)$ corresponds to the class of rotations of angle $\pi$ around axes orthogonal to the
axis of the drift dynamics. This is another scenario where our analysis applies.

The specific form of the interaction (Ising) is crucial  for the derivation of our results, and for the equivalence
of the two problems mentioned above. An interaction term of the form $\sigma_1 \otimes \sigma_2$, with arbitrary
$\sigma_1$ and $\sigma_2$, can be considered under the perspective of driving only one two-level system, the second
one representing an undesired disturbance. But in this case this problem is not equivalent to that of simultaneously
driving a pair of two-level systems with opposite drifts. Finally, more complicated interaction terms (e.g., the
Heisenberg interaction) are not compatible with our procedure.

In our analysis, the drift parameter $\omega_0$ and the control strength $\gamma$ can be arbitrarily varied,
and we have analytically derived the optimal control strategy and the corresponding time when $\frac{\gamma}{\omega_0}$
exceeds a threshold, which is approximatively $0.325$. For smaller control strength, numerical investigations are
needed.

We have found that the optimal control strategy for the SWAP operator (or equivalent) always requires $u_z = 0$,
both for nonsingular or singular arcs. Therefore, we can conclude that, for this class of operators, the scenario
with only two independent controls $u_x$ and $u_y$ is completely  equivalent to that considered in this work. The same
result has been found when investigating the analogous optimal control problem with only one two-level system~\cite{RR},
and it is a consequence of the symmetry of the problem.

Our results add to the growing literature on optimal control of quantum systems. The problem of time optimal
simultaneous control of two quantum bits in minimum time does not seem to have been considered earlier. One
notable exception is~\cite{Sugny}, where however the problem was set up for the state and not for the evolution
operator as here.



\appendix

\section{Numerical analysis supporting Conjecture 1}\label{app1}

In this appendix we provide numerical evidence that $L = 0$ is a necessary condition for the optimality of extremal
trajectories. Without loss of generality, we write the initial condition as $(b_x (0), 0)$, which,  with the
choice $E = \frac{\omega^2_0}{2}$, leads to
\begin{equation}\label{varth}
	\dot{b}_x (0) = \cos{\vartheta} \sqrt{1 - b_x (0)^2}, \qquad \dot{b}_y (0) = \sin{\vartheta} \sqrt{1 - b_x (0)^2}
\end{equation}
for some $\vartheta$ satisfying $0 \leqslant \vartheta \leqslant \pi$. We solve the costate
dynamics (\ref{syst3}), and from them we derive the corresponding extremal controls (\ref{optc}). Using the extremal controls,
we numerically solve Eq.s (\ref{schr2}) and derive the corresponding extremal  trajectories in the Lie group, $X_1 = X_1 (t, \vartheta, b_x (0))$
and $X_2 = X_2 (t, \vartheta, b_x (0))$, in the time interval $0 \leqslant t \leqslant T$. We then consider the following real functions:
\begin{eqnarray}\label{Fpm}
	F_+ &=& F_+ (t, \vartheta, b_x (0)) = \frac{1}{4} \Big[ \Big( {\rm Tr} (X_1 - X_2) \sigma_x \Big)^2 +
	\Big( {\rm Tr} (X_1 - X_2) \sigma_y\Big)^2 \Big], \nonumber \\
	F_- &=& F_- (t, \vartheta, b_x (0)) = \frac{1}{4} \Big[ \Big( {\rm Tr} (X_1 + X_2) \sigma_x \Big)^2 +
	\Big( {\rm Tr} (X_1 + X_2) \sigma_y\Big)^2 \Big],
\end{eqnarray}
satisfying $0 \leqslant F_{\pm} \leqslant 1$. Moreover, the function  $F_+$ reaches its maximum if and only if $X_1 = X_2$ matches
some operator equivalent to SWAP (according to Proposition~\ref{rem1}), and similarly for $F_-$, with $X_1 = - X_2$. Therefore
the optimal time $t_f$ we want to determine is the smallest time such that $F_+ = 1$ or $F_- = 1$, and it can be found by
gradually increasing  the range of numerical integration $[0,T]$.  For any choice of $\omega_0$ and $\gamma$ that we have considered, we have
always found that the smallest time is associated with $\vartheta = 0$ or $\pi$, or $b_x (0) = 0$ or $1$, which means $L = 0$.
For sake of completeness, some contour plots of $F_{\pm}$ are reported in Fig.s~\ref{sha1} and \ref{sha2}. We do not numerically
compute the minimum time $t_f$, because once we know that  $L = 0$ we can obtain an  analytical result.

This numerical analysis is fully consistent with the analytical expressions of nonsingular extremals presented in
Section~\ref{sec4}. With suitable choices of $T$, that is, by sufficiently increasing it, it is possible to investigate the structure of
sub-optimal extremal arcs, and, again, we find consistency with our analytical results (see Fig.~\ref{sha2}).
On one side our numerical computations
 motivate the analytical evaluation of extremal strategies; on the other, they provide an independent check of these results, since
they are fully consistent with them.


\section{Analysis of nonsingular extremals for $X_f = i \sigma_y$}\label{app2}

System (\ref{system}) admits analytical solutions for several values of $n$. We recall that a necessary
condition for all these extremals is $X_2 (t_f) = - X_1 (t_f)$.

The case $n = 0$ is the only possible extremal with only one switching time for $u_x$. We find that system
(\ref{system}) is  inconsistent,
with the only exception of the case when  $\gamma = \omega_0$. In this special case, it must be $\cos{\omega \tilde{\tau}} = 0$ and
$X_1 (t_f) = - X_f$, and we find $t_f = 2 \tilde{t} = \sqrt{2} \frac{\pi}{\gamma}$ and
$b_x (0)$ is unconstrained.

If $n = 1$, the extremal control strategies have $3$ switching times. We find that
\begin{equation}\label{sqsin}
    \sin{\omega \bar{\tau}} = \frac{\omega}{2 \omega_0} \sqrt{\frac{\gamma \pm \omega_0}{\gamma}}, \quad
    \sin{\omega \tilde{\tau}} = \frac{\omega}{2 \gamma} \sqrt{\frac{\omega_0 \mp \gamma}{\omega_0}},
\end{equation}
where the signs of the two terms follows from Proposition~\ref{remadd1}, which implies that $\sin{\omega \bar{\tau}} \geqslant 0$,
$\cos{\omega \bar{\tau}} \leqslant 0$, and $\sin{\omega \tilde{\tau}} \geqslant 0$. The upper signs correspond to
$X_1 (t_f) = X_f$, the lower signs to $X_1 (t_f) = - X_f$. Eq.s (\ref{sqsin}) are
consistent with (\ref{ttil}) and (\ref{tbar}) when
\begin{equation}\label{bx01}
    b_x (0) = \sqrt{\left( \frac{\omega_0}{\gamma} \right)^2 - \frac{\gamma \pm \omega_0}{3 \gamma \mp \omega_0}}.
\end{equation}
\begin{figure}[t]
  \includegraphics[width=15cm]{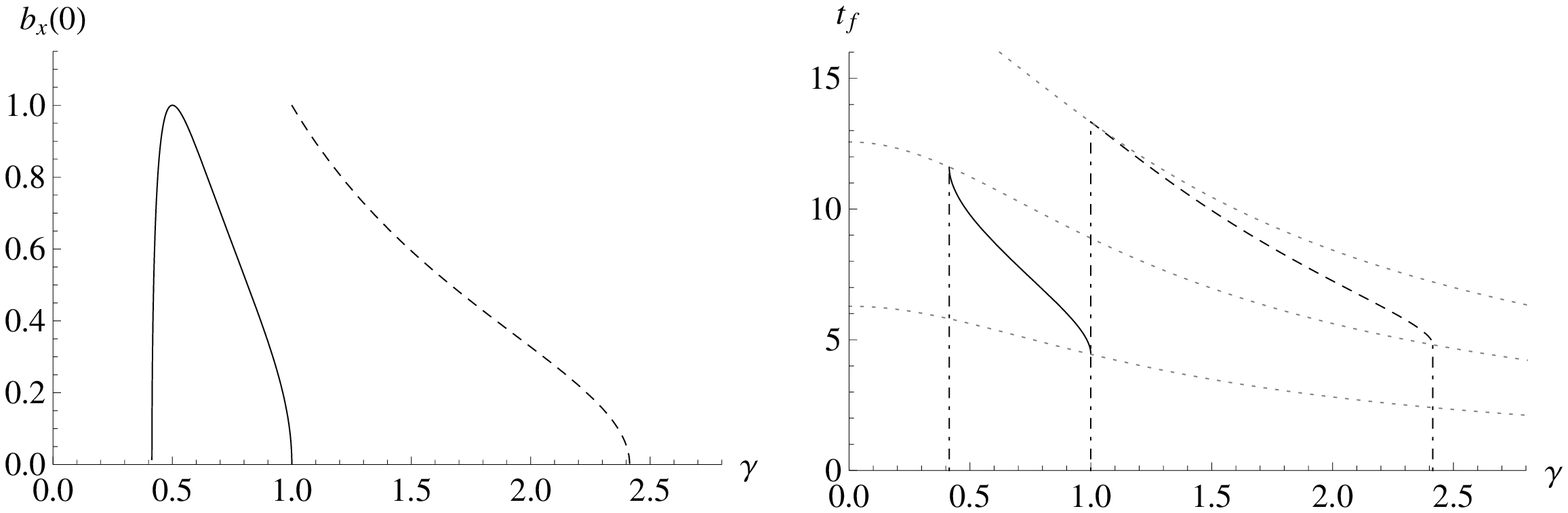}\\
  \caption{Plots of $b_x (0)$ and $t_f$ as functions of the control strength $\gamma$, with $\omega_0 = 1$, $X_f = i \sigma_y$
  and $n=1$. Dashed lines are associated with extremals which are certainly sub-optimal.}\label{sharday1}
\end{figure}
These solutions exist only when $\sqrt{2} - 1 \leqslant \frac{\gamma}{\omega_0} \leqslant 1$,
if we consider the upper sign, and when $1 \leqslant \frac{\gamma}{\omega_0} \leqslant \sqrt{2} + 1$ if we consider
the lower sign (otherwise the two functions in (\ref{sqsin}) are not defined, or $b_x (0) < 0$). The total time for the
transition is given by $t_f = 2 (\tilde{t} + \bar{t})$. For $X_1 (t_f) = X_f$  we find
\begin{equation}\label{finti1}
    t_f = \frac{4}{\sqrt{\omega_0^2 + \gamma^2}} \left( \pi - \arcsin{\frac{\omega}{2 \omega_0}
\sqrt{\frac{\gamma + \omega_0}{\gamma}}} + \arcsin{\frac{\omega}{2 \gamma}
\sqrt{\frac{\omega_0 - \gamma}{\omega_0}}}  \right).
\end{equation}
Similarly, for the transition $X_1 (t_f) = - X_f$ we obtain
\begin{equation}\label{finti1bis}
    t_f = \frac{4}{\sqrt{\omega_0^2 + \gamma^2}} \left( 2 \pi - \arcsin{\frac{\omega}{2 \omega_0}
\sqrt{\frac{\gamma - \omega_0}{\gamma}}} - \arcsin{\frac{\omega}{2 \gamma}
\sqrt{\frac{\omega_0 + \gamma}{\omega_0}}}  \right).
\end{equation}
In (\ref{finti1}) and (\ref{finti1bis}), we have used the fact  that the sign of $\cos{\omega \tilde{\tau}}$ is
determined by the third equation of (\ref{system}): it is positive in the first case, negative in
the second case. Plots of $b_x (0)$ and $t_f$ as functions of $\gamma$ are shown in Fig.~\ref{sharday1}.

If $n = 2$, the extremal control strategies have $5$ switching times. By expressing
$\cos{4 \alpha}$ and $\sin{4 \alpha}$ in terms of $\sin{\alpha}$ and $\cos{\alpha}$,
we rewrite system (\ref{system}) as
\begin{equation}\label{systemnew}
    \left\{ \begin{array}{l}
              1 - 8 \sin^2{\alpha} \left(1 - \sin^2{\alpha}\right) = \mp 2 \frac{\omega_0 \gamma}{\omega^2} \sin^2{\omega \tilde{\tau}} \\
               4 \frac{\gamma}{\omega_0} \sin^2{\alpha} \left(1 - 2 \sin^2{\alpha}\right) = \pm \left( 1 - 2 \left( \frac{\gamma}{\omega} \right)^2 \sin^2{\omega \tilde{\tau}} \right) \\
               4 \sin{\alpha} \left(1 - 2 \sin^2{\alpha}\right) \cos{\omega \bar{\tau}} = \mp \frac{\gamma}{\omega} \sin{2 \omega \tilde{\tau}}
            \end{array}
    \right..
\end{equation}
From the first and second equation we derive the biquadratic equation in $\sin{\alpha}$,
\begin{equation}\label{biqua}
    \sin^4{\alpha} - \frac{3}{4} \sin^2{\alpha} + \frac{\omega_0 \pm \gamma}{16 \gamma} = 0,
\end{equation}
which admits four solutions. If $X_1 (t_f) = X_f$, we obtain
\begin{equation}\label{quasol1}
    \sin{\alpha} = \frac{1}{2} \sqrt{\frac{3}{2} \pm \sqrt{\frac{5}{4} - \frac{\omega_0}{\gamma}}}.
\end{equation}
Conversely, if $X_1 (t_f) = - X_f$, we find
\begin{equation}\label{quasol2}
    \sin{\alpha} = \frac{1}{2} \sqrt{\frac{3}{2} \pm \sqrt{\frac{5}{4} + \frac{\omega_0}{\gamma}}}.
\end{equation}
\begin{figure}[t]
  \includegraphics[width=15cm]{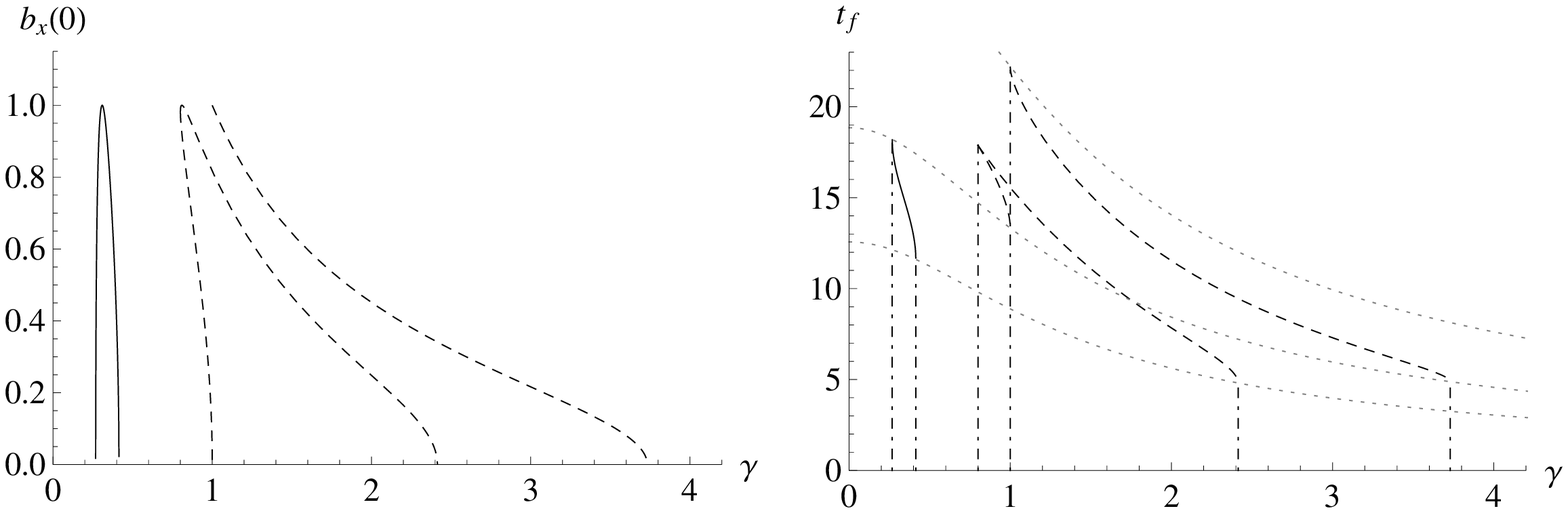}\\
  \caption{Plots of $b_x (0)$ and $t_f$ as functions of the control strength $\gamma$, with $\omega_0 = 1$, $X_f = i \sigma_y$
  and $n=2$. Dashed lines are associated with extremals which are certainly sub-optimal.}\label{sharday2}
\end{figure}
From Eq. (\ref{albe}) we can find $\sin{\omega \bar{\tau}}$, and moreover, from the first equation in (\ref{systemnew}),
by choosing the suitable signs, and rewriting the left hand side as $\cos{4 \alpha}$ for simplicity, we can evaluate
\begin{equation}\label{sinlast}
    \sin{\omega \tilde{\tau}} = \sqrt{\mp \frac{\omega^2}{2 \omega_0 \gamma} \cos{4 \alpha}}.
\end{equation}
These conditions are consistent with (\ref{ttil}) and (\ref{tbar}) as long as
\begin{equation}\label{bx03}
    b_x (0) = \sqrt{\frac{\omega_0^2 - \omega^2 \sin^2{\alpha}}{\gamma^2 \cos^2{\alpha}}}.
\end{equation}
Now, by requiring $0 \leqslant \sin{\alpha} \leqslant 1$ and $0 \leqslant b_x (0) \leqslant 1$, it is possible to
prove that all the solutions of (\ref{biqua}) are consistent, and to determine the associated ranges of $\frac{\gamma}{\omega_0}$,
which are not reported here because they are not needed for further developments. The final times associated to these four
extremals are given by $t_f = 2 \tilde{t} + 4 \bar{t}$, that is
\begin{equation}\label{finti1ter}
    t_f = \frac{2}{\sqrt{\omega_0^2 + \gamma^2}} \left( 4 \pi + 2 \arcsin{\sqrt{ \mp \frac{\omega^2}{2 \omega_0 \gamma} \cos{4 \alpha}}} - 4 \arcsin{\left(\frac{\omega}{\omega_0} \sin{\alpha}\right)} \right).
\end{equation}
Plots of $b_x (0)$ and $t_f$ for these four extremal trajectories are shown in Fig.~\ref{sharday2}. Our development shows that the only
extremal which can contribute to optimal solutions corresponds to the solution with positive sign in Eq. (\ref{quasol2}),
which is associated with $X_1 (t_f) = - X_f$.

\section{Analysis of nonsingular extremals for $X_f = i \sigma_x$}\label{app3}

We compute the analytical solution of system (\ref{system2}) in the simplest cases.
We recall that in this case it must be $X_1 (t_f) = X_2 (t_f) = X_f$.

\begin{figure}[t]
  \includegraphics[width=15cm]{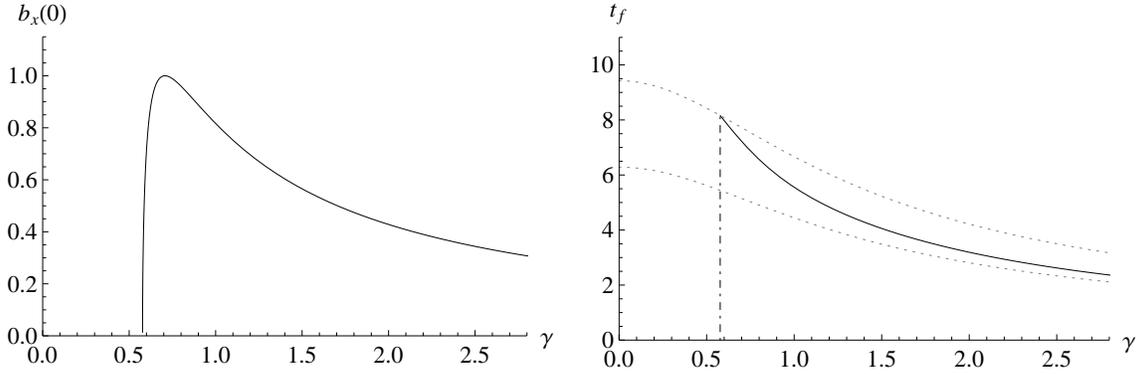}\\
  \caption{Plots of $b_x (0)$ and $t_f$ as functions of the control strength $\gamma$, with $\omega_0 = 1$, $X_f = i \sigma_x$
  and $n=0$. Dashed lines are associated with extremals which are certainly sub-optimal.}\label{sharday3}
\end{figure}

If $n = 0$, the extremal control strategies have $2$ switching times. By considering the definitions (\ref{albe}), the system is rewritten as
\begin{equation}\label{system3}
    \left\{ \begin{array}{l}
              \cos{\omega \bar{\tau}} = \mp \frac{\gamma}{\omega} \sin{2 \omega \tilde{\tau}} \\
              \frac{\gamma}{\omega} \sin{\omega \bar{\tau}} = \pm \left(1 - 2 \left( \frac{\gamma}{\omega} \right)^2
              \sin^2{\omega \tilde{\tau}} \right) \\
              \frac{\omega_0}{\omega} \sin{\omega \bar{\tau}} = \pm 2 \frac{\omega_0 \gamma}{\omega^2}  \sin^2{\omega \tilde{\tau}}
            \end{array}
    \right..
\end{equation}
Consider the case of upper signs in (\ref{system3}), that is, final operator $X_1 (t_f) = X_f$.
Since $\omega_0 \ne 0$, we find $\sin{\omega \tilde{\tau}} = \pm \frac{\omega}{2 \gamma}$ and
$\sin{\omega \bar{\tau}} = \frac{\omega}{2 \gamma}$, which are well defined only when $\frac{\omega}{2 \gamma} \leqslant 1$,
or $\frac{\gamma}{\omega_0} \geqslant \frac{1}{\sqrt{3}}$. Moreover, from $\sin{\omega \tilde{\tau}} = \pm \sin{\omega \bar{\tau}}$
and the first equation in (\ref{system3}), we find that $\cos{\omega \tilde{\tau}} = \mp \cos{\omega \bar{\tau}}$. We conclude that
$\cos{\omega \tilde{t}} = \cos{\omega \bar{t}}$ and $\sin{\omega \tilde{t}} = - \sin{\omega \bar{t}}$. These are constraints
on the switching times, which must be consistent with the expressions (\ref{ttil}) and (\ref{tbar}). By using the explicit expressions
on $A_0$, $B_0$ and $C_0$, we find that the initial condition of the costate must satisfy
\begin{equation}\label{bx02}
    b_x (0) = \frac{\omega_0}{\gamma} \sqrt{\frac{(\omega_0^2 - 2 \gamma^2)^2 - \gamma^4}{(\omega_0^2 - 2 \gamma^2)^2 - \omega_0^2 \gamma^2}}.
\end{equation}
Therefore, this extremal is well defined when $\gamma \geqslant \frac{1}{\sqrt{3}} \omega_0$, and the total time
for the transition is given by $t_f = 2 \tilde{t} + \bar{t}$, or more explicitly
\begin{equation}\label{finti2}
    t_f = \frac{2}{\sqrt{\omega_0^2 + \gamma^2}} \left( \pi + \arctan{\sqrt{\frac{\omega_0^2 + \gamma^2}
    {3 \gamma^2 - \omega_0^2}}} \right).
\end{equation}
The case with final operator $X_1 (t_f) = - X_f$ is obtained by replacing $\tau$ with $\tau + \pi$ in the previous formulas.
However, the result $\sin{\omega \bar{\tau}} = - \frac{\omega}{2 \gamma} < 0$ is inconsistent with the conditions
expressed in Proposition~\ref{remadd1}, therefore there are no extremal trajectories in this case. See Fig.~\ref{sharday3}
for plots of $b_x (0)$ and $t_f$ in this case.

If $n = 1$, the extremal control strategies have $4$ switching times, and the system (\ref{system2}) becomes
\begin{equation}\label{system3bis}
    \left\{ \begin{array}{l}
              \left( 4 \sin^2{\alpha} - 1 \right) \cos{\omega \bar{\tau}} = \pm \frac{\gamma}{\omega} \sin{2 \omega \tilde{\tau}} \\
              \frac{\gamma}{\omega_0} \left( 1 - 4 \sin^2{\alpha} \right) \sin{\alpha} = \pm \left(1 - 2 \left( \frac{\gamma}{\omega} \right)^2
              \sin^2{\omega \tilde{\tau}} \right) \\
             \sin{\alpha} \left( 3 - 4 \sin^2{\alpha} \right) = \pm 2 \frac{\omega_0 \gamma}{\omega^2}  \sin^2{\omega \tilde{\tau}}
            \end{array}
    \right..
\end{equation}
From the second and third equation of this system we get a depressed cubic equation in $\sin{\alpha}$,
\begin{equation}\label{depr}
    \sin^3{\alpha} + p \sin{\alpha} + q = 0,
\end{equation}
where
\begin{equation}\label{pq}
    p = - \frac{1}{2}, \qquad q = \pm \frac{\omega_0}{8 \gamma}.
\end{equation}
This equation can be solved by using the Cardano's method. By defining $u$ and $v$ such that $u + v = \sin{\alpha}$,
we find $u^3 + v^3 = - q$ and $3 u v = - p$, we find that both $u^3$ and $v^3$ satisfy the equation
\begin{equation}\label{z}
    z^2 + q z - \Big( \frac{p}{3} \Big)^3 = 0.
\end{equation}
Then we solve this quadratic equation and get $u^3$ and $v^3$, extract the cubic roots, and impose the further requirement
$3 u v = - p$. We are interested in real roots of the original cubic equation, and their number depend on the sign of
the discriminant $\Delta = -27 q^2 - 4 p^3$. If $\Delta < 0$, there is only one real root; if $\Delta = 0$ there are three real
roots with degeneration, and finally if $\Delta > 0$ there are three distinct real roots. By considering the values of $p$
and $q$ in (\ref{pq}), we conclude that, if $\frac{\gamma}{\omega_0} < \frac{3}{4} \sqrt{\frac{3}{2}}$, the only real root is
given by
\begin{equation}\label{reroo}
	\sin{\alpha} = \sqrt[3]{-\frac{q}{2} + \sqrt{\Big(\frac{q}{2}\Big)^2 + \Big(\frac{p}{3}\Big)^3}} +
	\sqrt[3]{-\frac{q}{2} - \sqrt{\Big(\frac{q}{2}\Big)^2 + \Big(\frac{p}{3}\Big)^3}},
\end{equation}
where we consider the real cubic root.
Since it must be $\sin{\alpha} \geqslant 0$, we require $q \leqslant 0$, and then the only possibility is $X_1 (t_f) = - X_f$,
corresponding to the lower sign in (\ref{system3bis}) and (\ref{pq}). If $\frac{\gamma}{\omega_0} \geqslant \frac{3}{4} \sqrt{\frac{3}{2}}$,
all the roots are real, and their general expressions are
\begin{equation}\label{reroobis}
	(\sin{\alpha})_k = 2 \sqrt{- \frac{p}{3}} \cos{\Big(\delta + \frac{2 k}{3} \pi}\Big),
\end{equation}
with $k = 0, 1, 2$, and
\begin{equation}\label{delta}
	\delta = - \frac{1}{3} \arctan{\frac{1}{3 q}\sqrt{\frac{\Delta}{3}}}.
\end{equation}
In the given range of parameters, $\sin{3 \alpha}$ is positive for $k = 0$ and negative for $k = 1, 2$.
Therefore, in the first case we have to consider the upper signs in (\ref{system3bis}) and (\ref{pq}), leading to $X_1 (t_f) = X_f$,
and the lower signs otherwise, with $X_1 (t_f) = - X_f$.

\begin{figure}[t]
  \includegraphics[width=15cm]{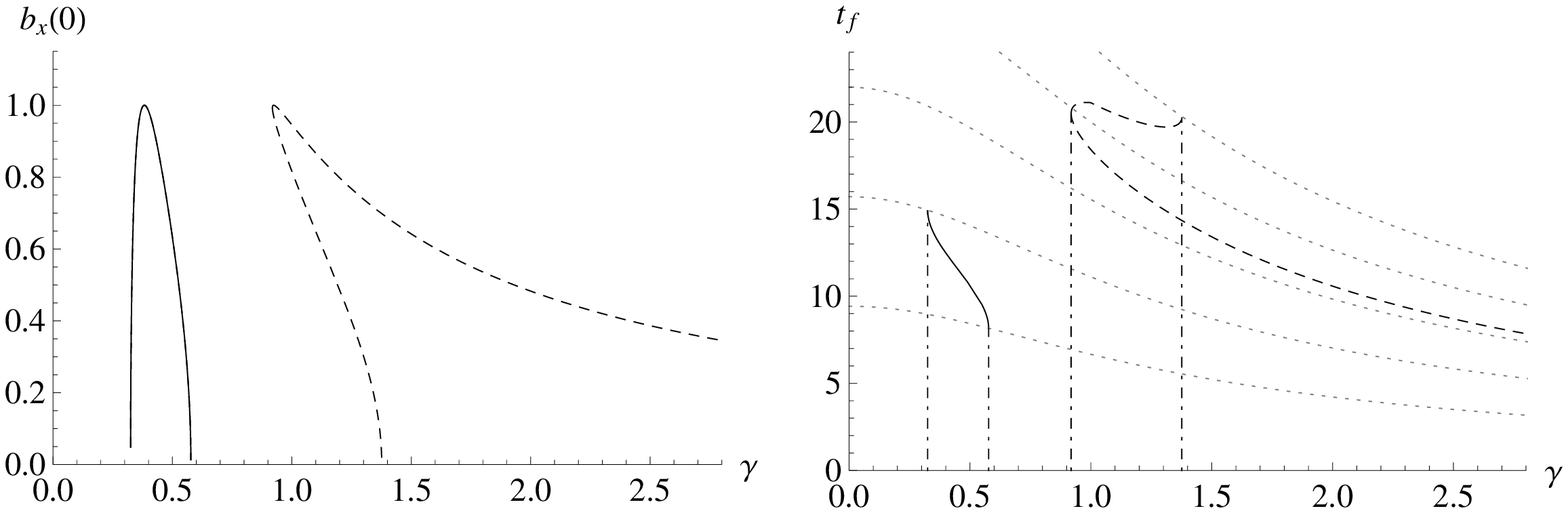}\\
  \caption{Plots of $b_x (0)$ and $t_f$ as functions of the control strength $\gamma$, with $\omega_0 = 1$, $X_f = i \sigma_x$
  and $n=1$. Dashed lines are associated with extremals which are certainly sub-optimal.}\label{sharday4}
\end{figure}

The requirements $0 \leqslant \sin{\alpha} \leqslant 1$ and $0 \leqslant \sin{\omega \bar{\tau}} \leqslant 1$ further restrict
the range of admissible values for $\frac{\gamma}{\omega_0}$. Their explicit expression are cumbersome and unnecessary;
we only mention that extremal trajectories associated to the case $k = 0$ in (\ref{reroobis}) are impossible because of the
constraint $\sin{\omega \bar{\tau}} \leqslant 1$. Therefore, $X_1 (t_f) = - X_f$ is the only possible final operation.
By solving the last equation in (\ref{system3bis}) we find
\begin{equation}\label{reroo2}
	\sin{\omega \tilde{\tau}} = \sqrt{- \frac{\omega^2}{2 \omega_0 \gamma} \sin{3 \alpha}},
\end{equation}
which is well defined because $\sin{3 \alpha} \leqslant 0$.
The conditions (\ref{reroo}) and (\ref{reroo2}) are consistent with (\ref{ttil}) and (\ref{tbar}) as long as (\ref{bx03}) is satisfied.
The total time for the transition on this extremal trajectory is given by $t_f = 2 \tilde{t} + 3 \bar{t}$, that is
\begin{equation}\label{finiti2bis}
	t_f = \frac{2}{\sqrt{\omega_0^2 + \gamma^2}} \left( 3 \pi + 2 \arcsin{\sqrt{- \frac{\omega^2}{2 \omega_0 \gamma} \sin{3 \alpha}}} - 3 \arcsin{\left(\frac{\omega}{\omega_0} \sin{\alpha}\right)} \right).
\end{equation}
Fig.~\ref{sharday4} shows plots of $b_x (0)$ and $t_f$ for this class of extremals. According to the analysis of Section~\ref{sec6},
the only extremal which can be optimal is the one characterized by (\ref{reroo}).


\section{Analysis of singular extremals}\label{app4}

Considering the two terms in (\ref{singu}), and assuming that $B_0 \leqslant 0$, we can be recast the final result in a rather compact form as
\begin{eqnarray}\label{X1sing}
	X_1 (t) &=& - \Big( \sin{\eta} \sin{(\xi_+ + \eta)} - \cos{\varphi} \cos{(\xi_+ + \varphi)}  \Big) I + \nonumber \\
		&-& \frac{i}{2} \Big((1 - \cos{2 \varphi}) \cos{(\xi_+ + \eta)} + \sin{2 \varphi} \sin{(\xi_+ + \eta)}  \Big) \sigma_x + \nonumber \\
		&-& \frac{i}{2} \Big((1 + \cos{2 \varphi}) \sin{(\xi_+ + \eta)} + \sin{2 \varphi} \cos{(\xi_+ + \eta)}  \Big) \sigma_y + \nonumber \\
		&-& i \Big( \sin{\eta} \cos{(\xi_+ + \eta)} - \sin{\varphi} \cos{(\xi_+ + \varphi)}  \Big) \sigma_z
\end{eqnarray}
and
\begin{eqnarray}\label{X2sing}
	X_2 (t) &=& + \Big( \sin{\eta} \sin{(\xi_- - \eta)} + \cos{\varphi} \cos{(\xi_- + \varphi)}  \Big) I + \nonumber \\
		&-& \frac{i}{2} \Big((1 - \cos{2 \varphi}) \cos{(\xi_- - \eta)} + \sin{2 \varphi} \sin{(\xi_- - \eta)}  \Big) \sigma_x + \nonumber \\
		&-& \frac{i}{2} \Big((1 + \cos{2 \varphi}) \sin{(\xi_- - \eta)} + \sin{2 \varphi} \cos{(\xi_- - \eta)}  \Big) \sigma_y + \nonumber \\
		&+& i \Big( \sin{\eta} \cos{(\xi_- - \eta)} + \sin{\varphi} \cos{(\xi_- + \varphi)}  \Big) \sigma_z
\end{eqnarray}
where we have defined the two functions
\begin{equation}\label{parxi}
	\xi_{\pm} = \xi_{\pm} (t^{\prime}) = \frac{1}{2} \left( \int_0^{t^{\prime}} u_z (s) ds \pm \omega_0 t^{\prime} \right)
\end{equation}
and $\eta \in (0, \frac{\pi}{2})$ satisfies
\begin{equation}\label{pareta}
	\sin{\eta} = \frac{\omega_0}{\gamma}, \qquad \cos{\eta} = \frac{1}{\gamma} \sqrt{\gamma^2 - \omega_0^2}.
\end{equation}
Since $u_z$ enters the problem only through the integral in (\ref{parxi}), without loss of generality we can assume it is constant on the singular arc. Therefore $\xi_{\pm} = \frac{1}{2} (u_z \pm \omega_0) t^{\prime}$. The free parameters in (\ref{X1sing}) and
(\ref{X2sing}) are $\varphi$ and $\xi_{\pm}$ (or, which is the same, $\varphi$, $u_z$ and $t^{\prime}$).
Target operators equivalent to SWAP have the form
\begin{equation}\label{swapeq}
	X_f = i e^{i \varphi S_z} \sigma_y e^{- i \varphi S_z} = i (\sin{\varphi} \sigma_x + \cos{\varphi} \sigma_y),
\end{equation}
and, by requiring $X_1 = \pm X_f$ we find the system
\begin{equation}\label{system4}
    \left\{ \begin{array}{l}
              \sin{\eta} \sin{(\xi_+ + \eta)} - \cos{\varphi} \cos{(\xi_+ + \varphi)}  = 0 \\
              (1 - \cos{2 \varphi}) \cos{(\xi_+ + \eta)} + \sin{2 \varphi} \sin{(\xi_+ + \eta)} =  \mp 2 \sin{\varphi} \\
              (1 + \cos{2 \varphi}) \sin{(\xi_+ + \eta)} + \sin{2 \varphi} \cos{(\xi_+ + \eta)} = \mp 2 \cos{\varphi} \\
              \sin{\eta} \cos{(\xi_+ + \eta)} - \sin{\varphi} \cos{(\xi_+ + \varphi)} = 0
            \end{array}
    \right.
\end{equation}
From the first and fourth equations we find that $\cos{(\xi_+ + \varphi)} = \pm \sin{\eta}$, and consequently
$\sin(\xi_+ + \eta) = \pm \cos{\varphi}$ and $\cos(\xi_+ + \eta) = \pm \sin{\varphi}$. We observe that the second and
third equations in (\ref{system4}) reduce to identities. Finally, the
solution of (\ref{system4}) is expressed as
\begin{equation}\label{solxi+}
    \cos{\xi_+} = \pm \sin{(\varphi + \eta)}, \qquad \sin{\xi_+} = \pm \cos{(\varphi + \eta)}.
\end{equation}
For solving $X_2 = \pm X_f$, we follow the same steps. We find that
\begin{equation}\label{solxi-}
    \cos{\xi_-} = \mp \sin{(\varphi - \eta)}, \qquad \sin{\xi_-} = \mp \cos{(\varphi - \eta)}.
\end{equation}
Since the signs in (\ref{solxi+}) and (\ref{solxi-}) are opposite, the only consistent solutions must
satisfy $X_1 = - X_2 = X_f$, with $X_f$ as in (\ref{swapeq}) for some $\varphi$. By combining Eq.s
(\ref{solxi+}) and (\ref{solxi-}), we find that $u_z t^{\prime} = 2 \pi - 2 \varphi + 4 k \pi$ or
$u_z t^{\prime} = 4 \pi - 2 \varphi + 4 k \pi$ for some integer $k$, and $\omega_0 t^{\prime} =
\pi - 2 \eta + 4 l \pi$ or $\omega_0 t^{\prime} = 3 \pi - 2 \eta + 4 l \pi$ for some integer $l \geqslant 0$.
We are interested in the solution with minimum time $t^{\prime}$, with the additional constraint
that $\vert u_z \vert \leqslant \gamma$. This minimum is obtained when $\varphi = 0$, $k = -1$ and
$l = 0$, that is, $\omega_0 t^{\prime} = \pi - 2 \eta$ and $u_z = 0$. This means that, on singular arcs on extremal trajectories, the evolution is given by the drift term.

Summing up, the total time for the transition $X \rightarrow X_f = i \sigma_y$ is given by
$t_f = 2 \tilde{t} + t^{\prime}$, where $\tilde{t}$ is determined by (\ref{cosi}). The explicit
expression in terms of $\omega_0$ and $\gamma$ is
\begin{equation}\label{finti3}
    t_f = \frac{2}{\sqrt{\omega_0^2 + \gamma^2}} \left(\pi - \arccos{\left( \frac{\omega_0}
    {\gamma}\right)^2 } \right) + \frac{1}{\omega_0} \left( \pi - 2 \arcsin{\frac{\omega_0}{\gamma}} \right).
\end{equation}

When $B_0 > 0$, the analysis is completely analogous. The dynamics in the Lie group is described by the following equations:
\begin{eqnarray}\label{X1sing2}
	X_1 (t) &=& + \Big( \sin{\eta} \sin{(\xi_+ - \eta)} + \cos{\varphi} \cos{(\xi_+ + \varphi)}  \Big) I + \nonumber \\
		&+& \frac{i}{2} \Big((1 - \cos{2 \varphi}) \cos{(\xi_+ - \eta)} + \sin{2 \varphi} \sin{(\xi_+ - \eta)}  \Big) \sigma_x + \nonumber \\
		&+& \frac{i}{2} \Big((1 + \cos{2 \varphi}) \sin{(\xi_+ - \eta)} + \sin{2 \varphi} \cos{(\xi_+ - \eta)}  \Big) \sigma_y + \nonumber \\
		&+& i \Big( \sin{\eta} \cos{(\xi_+ - \eta)} + \sin{\varphi} \cos{(\xi_+ + \varphi)}  \Big) \sigma_z
\end{eqnarray}
and
\begin{eqnarray}\label{X2sing2}
	X_2 (t) &=& - \Big( \sin{\eta} \sin{(\xi_- + \eta)} - \cos{\varphi} \cos{(\xi_- + \varphi)}  \Big) I + \nonumber \\
		&+& \frac{i}{2} \Big((1 - \cos{2 \varphi}) \cos{(\xi_- + \eta)} + \sin{2 \varphi} \sin{(\xi_- + \eta)}  \Big) \sigma_x + \nonumber \\
		&+& \frac{i}{2} \Big((1 + \cos{2 \varphi}) \sin{(\xi_- + \eta)} + \sin{2 \varphi} \cos{(\xi_- + \eta)}  \Big) \sigma_y + \nonumber \\
		&-& i \Big( \sin{\eta} \cos{(\xi_- + \eta)} - \sin{\varphi} \cos{(\xi_- + \varphi)}  \Big) \sigma_z.
\end{eqnarray}
As before, the only consistent requirement is $X_1 = - X_2 = X_f$, and the solution is given by
\begin{equation}\label{solxi+bis}
    \cos{\xi_-} = \pm \sin{(\varphi + \eta)}, \qquad \sin{\xi_-} = \pm \cos{(\varphi + \eta)}.
\end{equation}
For solving $X_2 = \pm X_f$, we follow the same steps. We find that
\begin{equation}\label{solxi-bis}
    \cos{\xi_+} = \mp \sin{(\varphi - \eta)}, \qquad \sin{\xi_+} = \mp \cos{(\varphi - \eta)}.
\end{equation}
By combining these equations, we find that $u_z t^{\prime} = 2 \pi - 2 \varphi + 4 k \pi$ or
$u_z t^{\prime} = 4 \pi - 2 \varphi + 4 k \pi$ for some integer $k$, and $\omega_0 t^{\prime} =
\pi + 2 \eta + 4 l \pi$ or $\omega_0 t^{\prime} = 3 \pi + 2 \eta + 4 l \pi$ for some integer $l \geqslant 0$.
These extremals are certainly not optimal, as both $\tilde{t}$ and $t^{\prime}$ are larger than the corresponding
quantities derived in the case $B_0 \leqslant 0$.


\begin{figure}[p]
  \includegraphics[width=13cm]{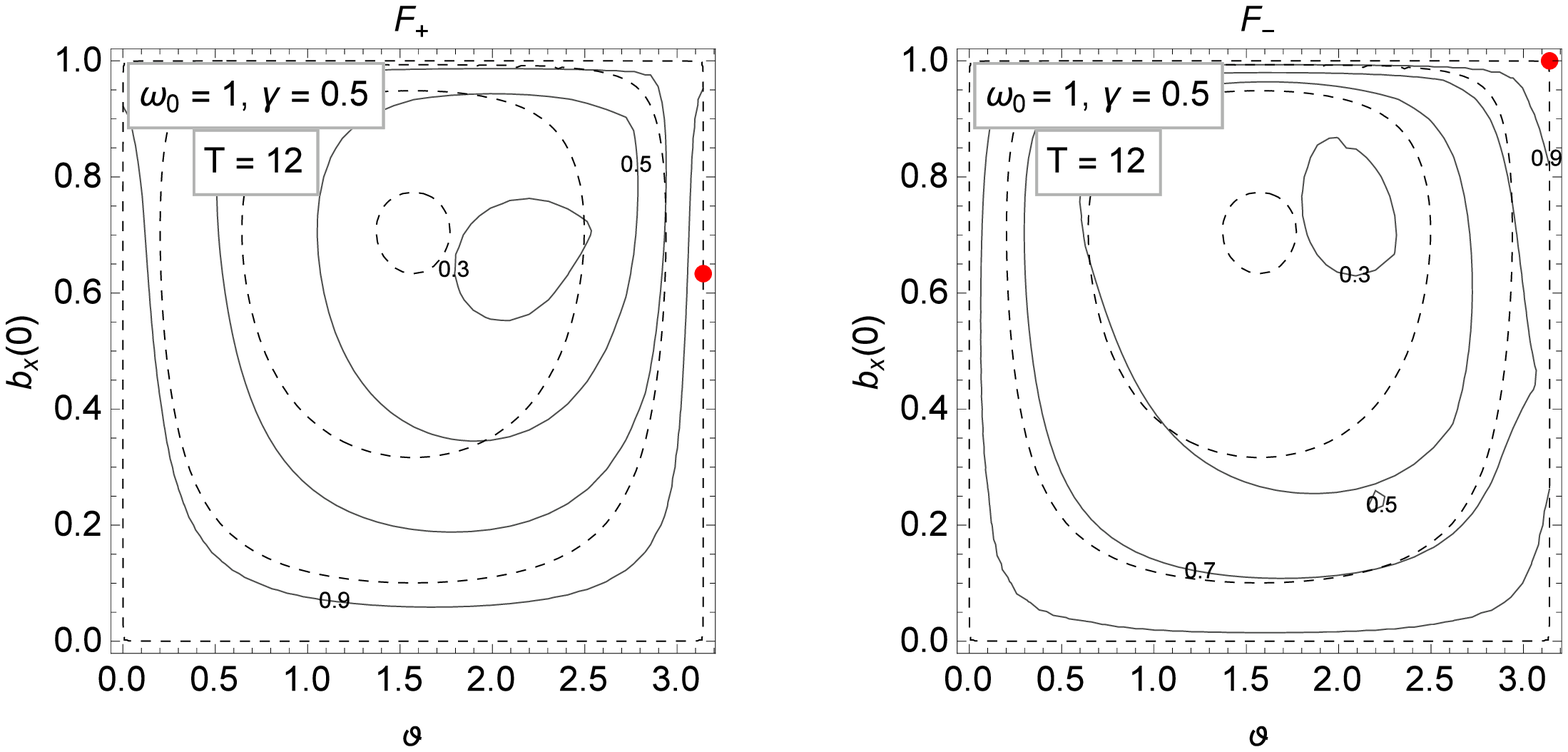}\\
\end{figure}
\begin{figure}
  \includegraphics[width=13cm]{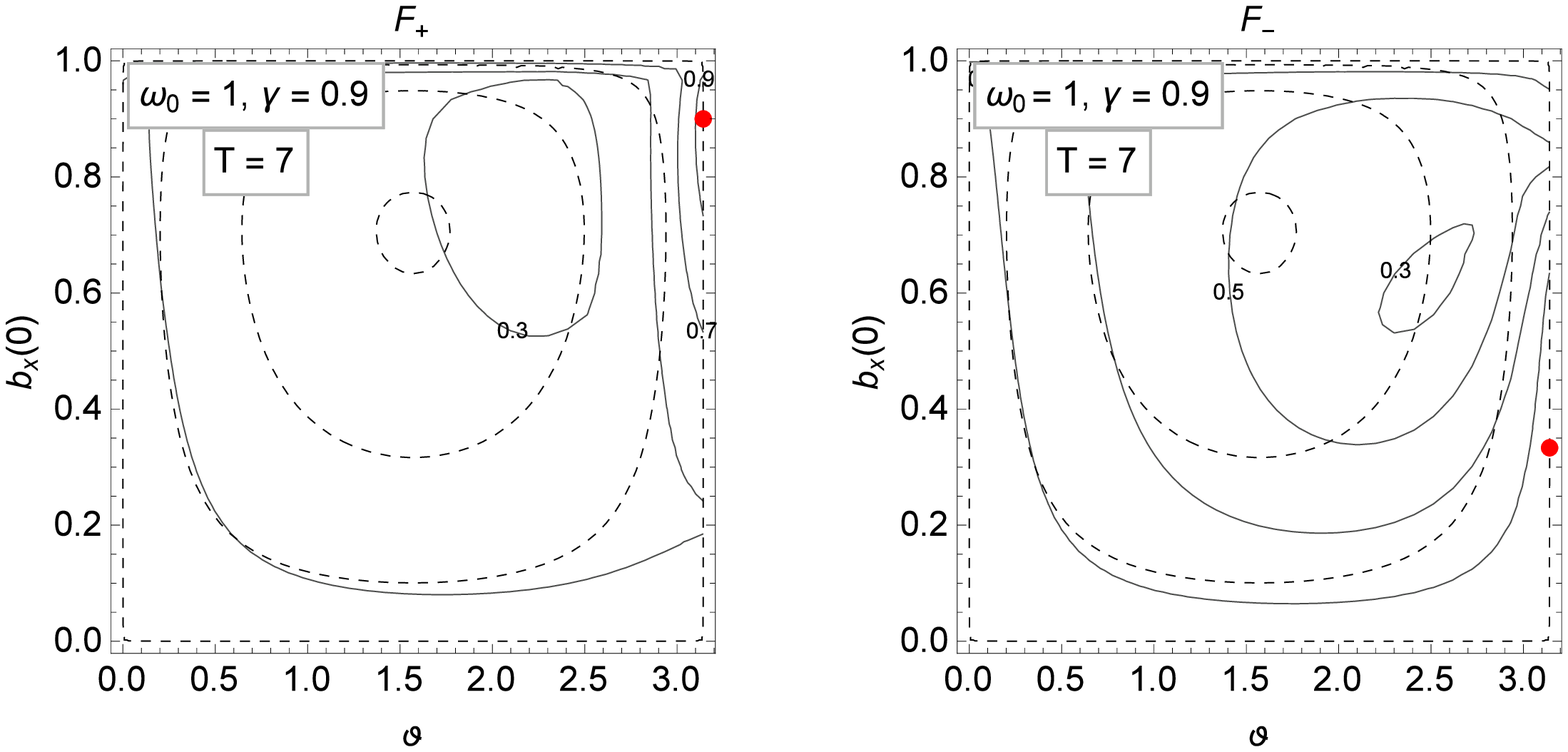}\\
\end{figure}
\begin{figure}
  \includegraphics[width=13cm]{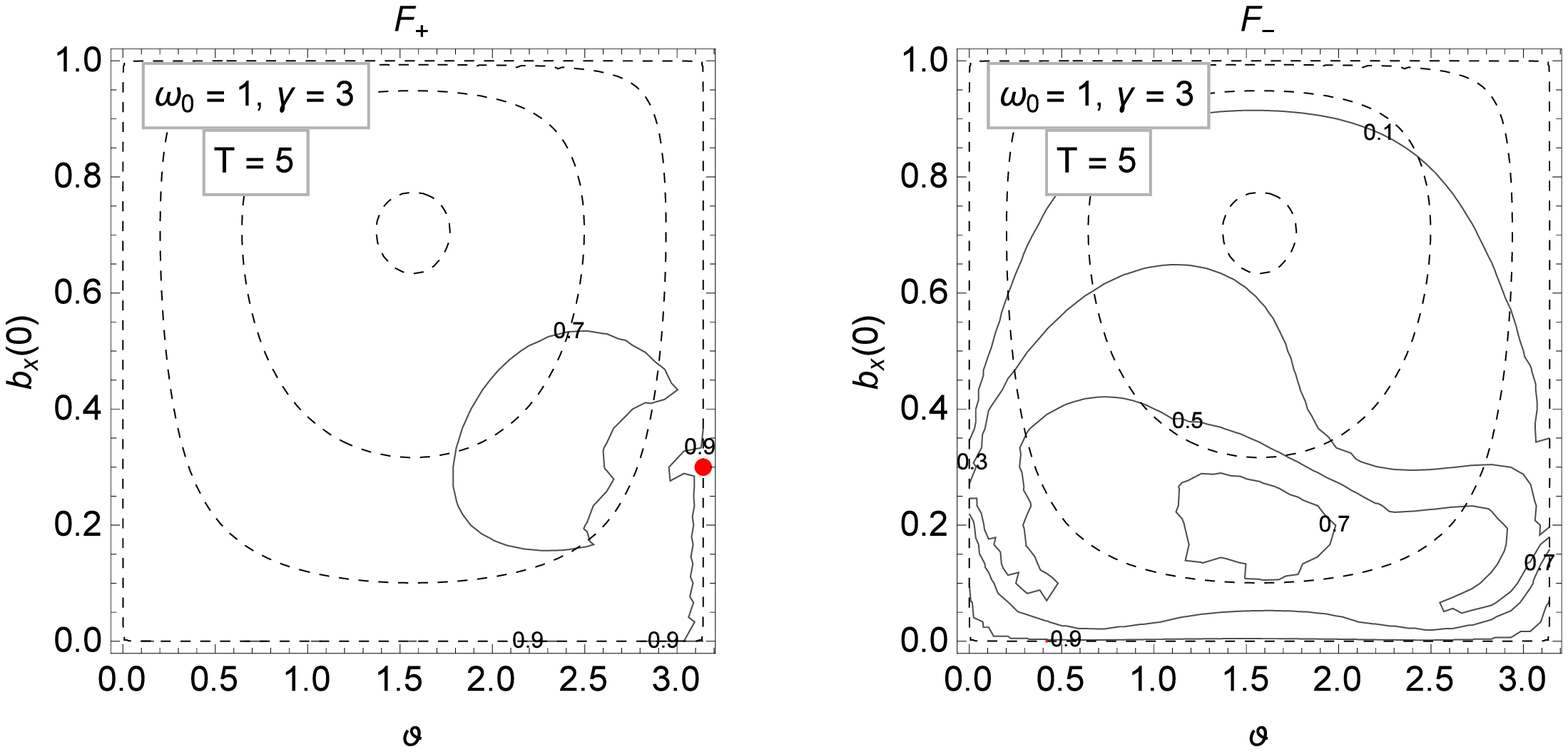}\\
  \caption{Contour plots of the functions $F_+$ and $F_-$ for selected values of $\omega_0$, $\gamma$ and $T$ (in the inset).
  The dashed lines represent points with a fixed value for $L$, decreasing towards the border of the plot, where $L = 0$.
  The red spots shows the values of $b_x (0)$ and $\vartheta$ such that $F_{\pm} = 1$ with $0 \leqslant t \leqslant T$.}\label{sha1}
\end{figure}
\begin{figure}
  \includegraphics[width=13cm]{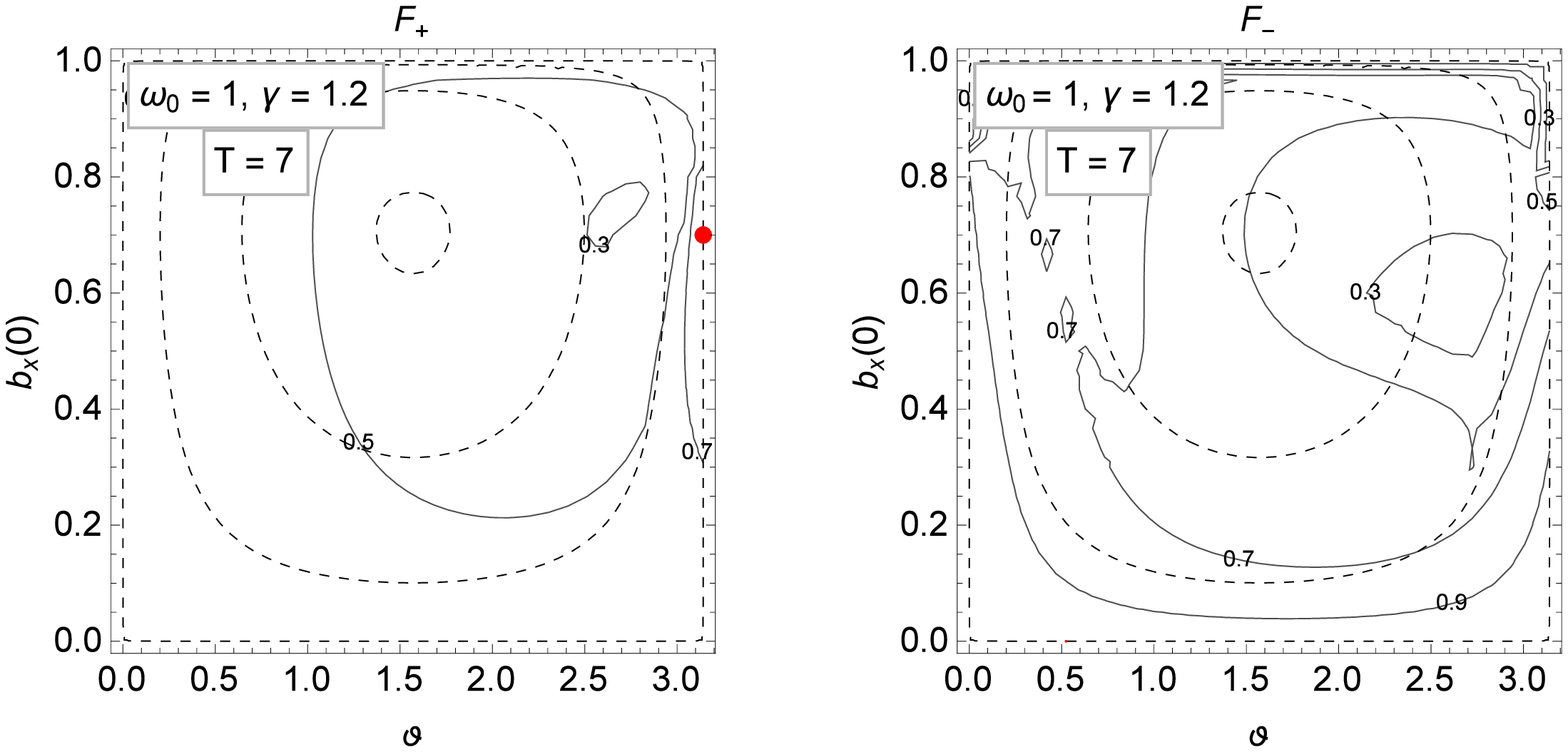}\\
\end{figure}
\begin{figure}
  \includegraphics[width=13cm]{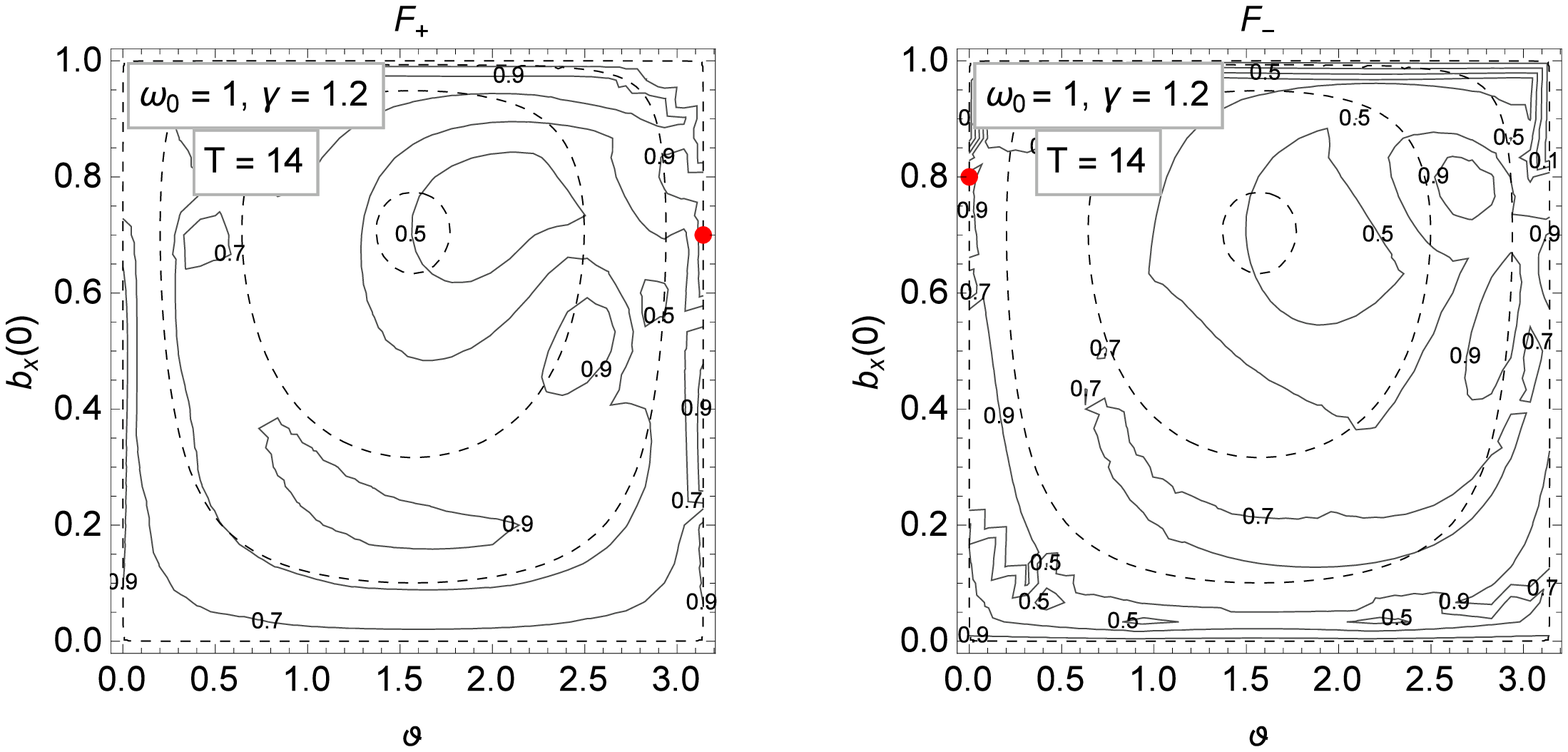}\\
  \caption{Contour plots of the functions $F_+$ and $F_-$ for selected values of $\omega_0$, $\gamma$ and $T$ (in the inset).
  The parameters in the two plots are the same, but $T$ which is increased. From the behavior of $F_-$, it is apparent the existence of
  a sub-optimal extremal.
  }\label{sha2}
\end{figure}


\end{document}